\documentclass[aps,pre,superscriptaddress,twocolumn]{revtex4-1}

\usepackage{amsmath,amssymb,amsfonts,amsthm,mathtools}
\usepackage{bm}

\usepackage{graphicx}
\usepackage{dcolumn}   % decimal-aligned columns
\usepackage{color}     % if you really need it; xcolor is nicer, but this is fine

\usepackage{textgreek} % upright greek in text

\usepackage[version=4]{mhchem}

\usepackage{tikz}
\usepackage[framemethod=tikz]{mdframed}

\usepackage{afterpage}
\usepackage{hyperref}
\hypersetup{
  colorlinks=true,
  linkcolor=blue,
  citecolor=blue,
  urlcolor=blue
}

\begin{document}

%\title{Mechanical principles of soft gel translocation under confinement}
%\title{How Does a Microgel Transport Through a Narrow Capillary?}
\title{Microgel Translocation Through Narrow Capillaries}
\author{Subhadip Biswas}
\email{subhadip.biswas@cnrs.fr}
\affiliation{Laboratoire de Physique des Solides, CNRS, Université Paris-Saclay, 91405 Orsay, France.}
\affiliation{Department of Physics and Astronomy, University of Sheffield, Sheffield S3 7RH, UK.}
\author{Buddhapriya Chakrabarti}
\email{buddho@ksu.edu}
% \altaffiliation{Department of Physics and Astronomy, University of Sheffield, Sheffield S3 7RH, UK.}

\affiliation{Department of Physics and Astronomy, University of Sheffield, Sheffield S3 7RH, UK.}
\affiliation{Department of Physics, Kansas State University, 1228 M. L. K. Jr. Drive, Manhattan, KS, 66506, USA.}

\date{\today}
\begin{abstract}
The transport of soft viscoelastic gels through confined geometries underlies critical processes in biomedical, biological, and industrial systems. Here, we examine the translocation of a spherical microgel through a narrow capillary whose diameter is smaller than the equilibrium gel size. Using coarse-grained molecular dynamics simulations in tandem with mean-field theory and mechanical analysis, we uncover a critical threshold diameter $d_c$ below which the microgel cannot enter, regardless of the applied pressure. This geometric limit emerges from the interplay between gel elasticity and its internal network connectivity, captured quantitatively by a graph-theoretic model. We construct a phase diagram in the parameter space of tube diameter $d$, applied force $f_g$, and gel stiffness $Y$ (Young’s modulus), which delineates the regimes of successful translocation and mechanical arrest. Under negligible wall friction, gel mobility scales with the applied force; however, beyond the cutoff parallel network topology, progressive densification in the constriction stalls the microgel. Our results reveal the mechanical and topological determinants of soft gel transport in confinement and provide predictive guidelines for engineering gel-based systems in microfluidics, drug delivery, and tissue-level filtration.
\end{abstract}

\maketitle 

\section{Introduction}
The translocation of polymers~\cite{p:Nelson1999,p:Park1996,p:Kardar2002,p:Kardar2004,p:Aksimentiev2004,p:Timp2005,p:Sakaue2007,p:Meller2008,b:Muthukumar2011,r:Wanunu2012,r:Palyulin2014,p:Sakaue2016}, and vesicles\cite{p:Muthukumar2016,p:Doi2018,p:Shi2019,p:Milchev2022,p:Doi2023,p:Auth2025} one and two-dimensional elastic objects through nanopores and nanochannels has been extensively studied over the past several decades. A unifying concept underlying these systems is that confinement into a space of lower dimensionality reduces the configurational entropy of polymers, giving rise to an entropic barrier that resists translocation. This barrier can be overcome by energetically favorable mechanisms such as interactions with the confining channel or the application of an external driving force. Within this framework, the free-energy landscape along a suitable reaction coordinate has been widely used to quantify translocation dynamics, with mean translocation times estimated using Kramers’ escape-rate theory \cite{kramers1940brownian}. 

An open question is whether these theoretical approaches, originally developed for lower-dimensional elastic objects, can be generalized to describe the translocation of liquids~\cite{p:Peng2023,p:Peng2023b} and three-dimensional soft materials, such as microgels~\cite{p:Baroud2023} or other deformable colloidal particles and cells~\cite{p:Guck2024}, through narrow channels. While conventional fluid mechanics~\cite{b:Landau1959,b:batchelor2000} can be readily applied to deal with the translocation of fluid drops, the translocation of soft solids is much more complex. Unlike polymers or membranes, these systems experience complex deformations that cannot be easily reduced to a one-dimensional reaction coordinate. Further, for gels whose dimensions exceed the constriction diameter, the deformation is highly non-affine—that is, the deformed configuration cannot be obtained from the undeformed one through a simple affine mapping governed by conventional stress–strain relations~\cite{p:Mackintosh2003,p:DiDonna2005,p:Yodh2012}. As a result, the notion of a single translocation coordinate or a well-defined free-energy barrier becomes ill-defined.

Understanding the translocation of three-dimensional viscoelastic gels, therefore, requires a theoretical framework that explicitly accounts for the coupled effects of microstructure, deformation mechanics, confinement, and flow. Such a framework is essential for describing soft particle and microgel transport in confined environments, which underpins many biological, biomedical, and industrial processes~\cite{p:Peng2023c,p:Viallat2020}. Representative examples include the passage of micron-sized hydrogels through pores under homeostatic pressure, mimicking renal filtration in which soft, deformable particles are cleared through narrow endothelial gaps~\cite{p:hendrickson2010}; the use of soft polymeric gel aggregates to occlude blood supply in tumour embolization or to treat aneurysms~\cite{p:carugo2012,p:laurent2004,p:osuga2012, p:freund2014}; and hydrogel-based drug delivery systems in which the gel serves as a transport vehicle for active pharmaceuticals~\cite{p:masoud2012, p:peppas1997,p:hoare2008,p:she2012}. Transport of soft objects also plays a role in enhanced oil recovery~\cite{p:lei2020, p:shi2011, p:son2016}. 

A common feature of these systems is that particles whose undeformed lateral dimensions exceed the channel width must deform to translocate under an imposed pressure difference~\cite{p:merker2011}. In this regime, particle–wall interactions dominate the transport dynamics. For sufficiently large particles, deformation may be insufficient, leading to stalling and blockage at the channel entrance~\cite{p:Baroud2023} with a concomitant buildup of upstream pressure. Despite the prevalence of such clogging events, direct measurements of the pressure drop across a blocked channel remain challenging.

During translocation through narrow constrictions, a microgel undergoes shape changes governed by its mechanical properties, flow conditions, entrance geometry, and size relative to the capillary dimensions~\cite{p:li2015}. Li et al. demonstrated a universal behavior for hydrogel entry and passage through corrugated narrow channels, showing quantitative agreement between experiments on biaxial gel deformation and an affine theoretical model for unentangled microgels under imposed strain~\cite{p:li2015, li2022fibrous}. 
Recent advances in microfluidic rheometry and controlled compression experiments have revealed how confinement and deformation can decouple structure from elasticity in soft materials, highlighting the need for predictive frameworks that link microscopic network topology to transport and mechanical response under extreme confinement \cite{milani2026decoupling, milani2026rheofluidics}. Motivated by the widespread use of gel-based systems in biomedical and industrial applications, extensive experimental and theoretical efforts have examined the transport of soft biomolecular objects through narrow capillaries~\cite{p:li2017, p:brooks2020, p:aguiar2019, p:lei2019, p:connell2019, p:khan2017, p:fai2017, p:nascimento2017, p:portnov2018, p:zhang2018, p:keith2020, p:buning2021, p:li2020, p:lesage2009, p:oevreeide2021, p:dai2010}.

Despite these advances, the conditions governing entry into a constricted capillary remain poorly understood. In particular, it is unclear how gel crosslinking controls the critical tube diameter $d_c$ below which a gel with a given crosslink distribution cannot enter the channel. Since crosslink topology directly determines the shear modulus of the network, a quantitative relation between elastic moduli and the pressure buildup required for translocation is still lacking. In the limit of high elasticity, the time-dependent deformation of elastic capsules traversing narrow capillaries has been investigated~\cite{p:rorai2015,p:kuriakose2013}. Complementary experiments on fluid permeation through polymer networks with varying microstructure have been used to quantify polymer–solvent friction~\cite{p:fujiyabu2017}. Such frictional coupling between the gel network and the solvent generates drag forces that critically influence gel motion and translocation through narrow confinements.

In this work, we investigate the entry and translocation of spherical microgels through narrow constrictions using coarse-grained molecular dynamics simulations complemented by semi-analytical theory. Beyond simple cylindrical or conical deformations, we observe a rich set of pressure-dependent gel morphologies controlled by network topology, elastic properties, and channel geometry. We show that the ability of a microgel to pass through a constriction under a given pressure is governed by its network connectivity rather than its equilibrium size or shape.

Building on this insight, we introduce a geometric criterion based on parallel network connectivity, formulated within undirected graph theory, to establish a lower bound on the constriction diameter below which a microgel with fixed connectivity cannot enter, independent of the applied pressure difference. This framework provides a direct link between physical parameters—such as channel geometry, confinement, and molecular interactions—and dynamical observables including flow rate, pressure buildup, and first-passage statistics. Our simulations further elucidate the mechanisms of gel entry into narrow capillaries and identify the conditions for successful versus failed translocation as functions of applied pressure, gel stiffness, and constriction size.

\begin{figure}[!h]
\centering
\includegraphics[width=\linewidth]{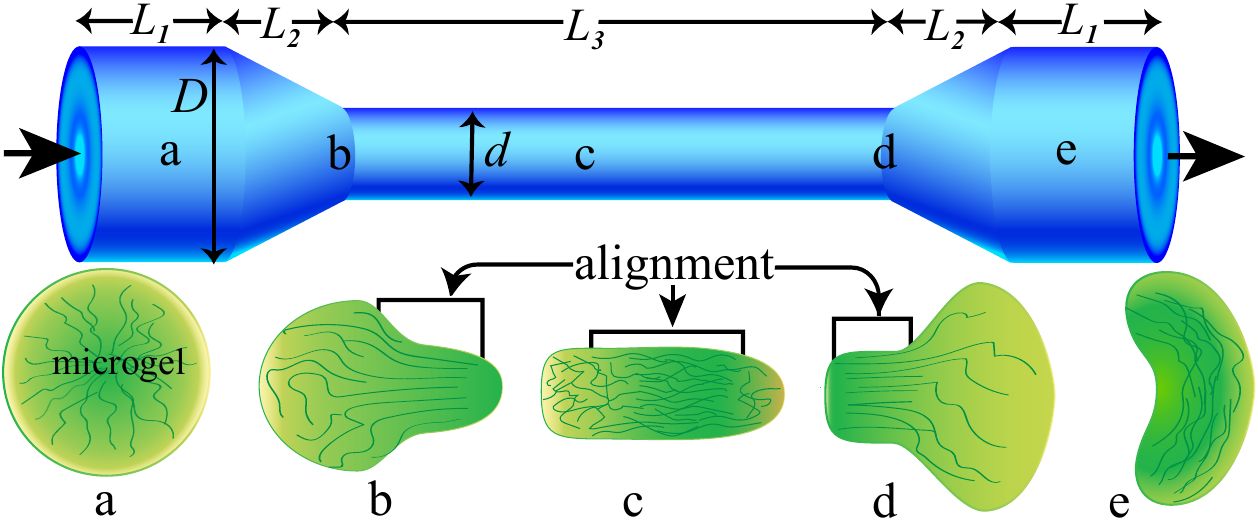}
\caption{Schematic of the corrugated tube geometry used in the simulations. The tube consists of a hollow cylinder of diameter $D>2R_{\mathrm{gel}}$ ({\bf a}), where $R_{\mathrm{gel}}$ is the undeformed microgel radius, followed by a conical frustum ({\bf b}) connecting to a constricted cylindrical segment of diameter $d<2R_{\mathrm{gel}}<D$ ({\bf c}). A mirror-symmetric conical frustum and wide cylinder are attached downstream ({\bf d} and {\bf e}), with the terminal wide cylinders ({\bf a} and {\bf e}) connected via periodic boundary conditions. Solvent flow is from left to right. The lower panel illustrates representative microgel conformations along the tube. An initially undeformed microgel resides in the wide section ({\bf a}) and encounters an entropic barrier at the constriction entrance ({\bf b}). Upon entry, the gel undergoes biaxial confinement and adopts an elongated cylindrical shape ({\bf c}). Downstream of the constriction, the gel expands upon release from confinement ({\bf d}) and subsequently forms a discoid shape due to solvent jetting from the narrow capillary ({\bf e}). Within the constricted region, network strands predominantly align with the tube axis.}
\label{fig:schematic_tube}%
\end{figure}

We construct a phase diagram delineating regimes in which microgels either translocate through or stall at the entrance of a narrow capillary. In the translocating regime, we quantify gel morphology, internal network organization (including nematic ordering), volume changes, and translocation times as functions of channel geometry and gel properties, and contrast these with the corresponding characteristics of stalled gels. We identify the confinement diameter $d$ and the applied pressure acting on gels of given stiffness as the key control parameters governing the motile--stalled transition. For a gel with shear modulus $G$ in a constriction of diameter $d$, entry requires the solvent force $f$ (equivalently, pressure $P$) to exceed a critical threshold $f_c$. A schematic of the channel geometry and representative gel conformations as it translocates through the channel, as obtained from simulations, is shown in Fig.~\ref{fig:schematic_tube}. We report passage times, volume transitions, solvent expulsion under confinement, spatially resolved network structure within corrugated channels, and orientational order parameters of network strands inside the capillary. These observables are quantitatively linked through a theoretical framework that accurately reproduces the coarse-grained molecular dynamics results and enables the prediction of microgel behavior under confinement.

The article is organized as follows. We first describe the simulation setup and methodology. We then establish criteria for successful gel translocation and present the resulting phase diagram, identifying the critical parameters governing passage through a constriction. To rationalize the transition, we introduce a quantitative framework based on a graph-theoretic approach and compute the parallel connectivity of the gel incorporating the geometric constraints imposed by the tube. We next analyze the deformation modes of translocating microgels, quantify strand orientational order within the constriction, and characterize the evolution of gel shape, volume, surface area, and solvent content during translocation. We also examine the properties of gels stalled at the constriction entrance. Finally, we summarize the main findings and outline possible directions for future work.

\section{Coarse-Grained Simulation of Microgel Translocation:}

The gel translocation simulations involve a two-step procedure. First, a crosslinked polymer microgel is generated and characterized, after which the equilibrated gel is inserted into the corrugated tube shown in Fig.~\ref{fig:schematic_tube}. The gelation protocol, along with the resulting morphological and mechanical properties, is well established~\cite{t:Biswas2022}. We consider two classes of microgels in the coarse-grained molecular dynamics (CGMD) simulations. The first consists of relatively small, stiff gels composed of $N=1120$ polymer strands with contour length $N_p=10$ and shear modulus $G \approx 0.3\,\epsilon/\sigma^3$, simulated in explicit solvent. The second class comprises larger gels with $N=1280$ polymer strands of contour length $N_p=20$, for which the cycle rank is varied to tune the shear modulus over the range $G \approx 0.003\,\epsilon/\sigma^3$ to $0.1\,\epsilon/\sigma^3$. Here, $\epsilon$ denotes the interaction energy between cross-linkers and $\sigma$ is the Lennard-Jones cutoff distance.

The corrugated channel is constructed from three connected three-dimensional segments: (i) a wide cylindrical section of diameter $D>2R_{\mathrm{gel}}$ and length $L_1$, where $R_{\mathrm{gel}}$ denotes the equilibrium swollen gel radius; (ii) a tapered region formed by a hollow conical frustum of length $L_2$, connecting diameters $D$ and $d$ ($D>d$); and (iii) a constricted cylindrical segment of diameter $d$ and length $L_3$. As shown in Fig.~\ref{fig:schematic_tube}, the geometry is reflection symmetric about the midpoint of the constricted region. Periodic boundary conditions applied at the ends of the wide cylinders generate an effectively infinite, periodically corrugated channel composed of alternating wide and narrow sections.

The channel geometry is specified by fixing the conical opening angle ($\sim 30^\circ$), the diameter of the wide cylinder $D$, and the constriction length $L_3$. For a given constriction diameter $d$ and entrance length $L_1$, the frustum length $L_2$ is determined geometrically. Gel translocation is investigated by varying the applied solvent force $f$ and the constriction diameter $d$. The channel is fully filled with solvent particles at a reduced density $\rho^* \approx 1$.

For the smaller gels ($N_p=10$), the channel parameters are as follows: entrance angle $30^\circ$, constriction length $L_3 = 140\sigma$, wide-cylinder radius $D/2 = 30.5\sigma > R_{\mathrm{gel}} \approx 21\sigma$, and constriction diameters in the range $d = 7.5\sigma$--$12.5\sigma$. The total channel length is fixed at $2(L_1+L_2)+L_3 = 300\sigma$. We note that typical vascular bifurcation angles are smaller than $30^\circ$~\cite{p:fanucci1988}. Unlike biological vessels, where radius variations are smooth and the effective entrance angle depends on the taper length~\cite{p:lesage2009, p:li2015}, the present model employs sharp transitions between cylindrical and conical segments to isolate geometric effects.

Polymer strands forming the crosslinked gel are modeled using the Kremer--Grest bead--spring framework with finitely extensible nonlinear elastic (FENE) bonds~\cite{p:Grest1986,p:Kremer1990},
\begin{eqnarray}
V_{\mathrm{FENE}}(r) = -\frac{k_b r_{\max}^2}{2}
\ln\!\left[1-\left(\frac{r}{r_{\max}}\right)^2\right],
\label{eq:fene}
\end{eqnarray}
where the bond stiffness is set to $k_b = 50\,\epsilon/\sigma^2$ and the maximum bond extension is $r_{\max}=1.5\sigma$. Simulations are performed in reduced Lennard--Jones (LJ) units, where $\sigma$ defines the unit of length, $\epsilon$ the unit of energy, $m$ the particle mass, and the unit of time is $\tau=\sqrt{m\sigma^2/\epsilon}$.

Excluded-volume interactions between nonbonded monomers are modeled using the purely repulsive Weeks--Chandler--Andersen (WCA) potential~\cite{p:Andersen1971},
\begin{eqnarray}
V_{\mathrm{rep}}(r)=
\begin{cases}
4\epsilon\!\left[\left(\frac{\sigma}{r}\right)^{12}
-\left(\frac{\sigma}{r}\right)^6+\frac{1}{4}\right], & r\le r_c,\\
0, & r>r_c,
\end{cases}
\label{eq:wca}
\end{eqnarray}
with cutoff distance $r_c=2^{1/6}\sigma$. No attractive interactions are included.

Chain stiffness can be increased by introducing a bending potential; however, to model soft polymer gels, all simulations are performed with fully flexible chains. To reduce computational cost, the confining channel is implemented via geometrical wall boundaries that interact with monomers through a repulsive LJ potential, rather than explicit wall particles.

An additional key control parameter is the intrinsic softness of the gel, characterized by its elastic modulus. While loosely crosslinked microgels readily translocate through narrow channels, stiffer gels of comparable size may stall at the entrance. We therefore prepare gels spanning a broad range of elastic moduli. Because the gel state (mobile or stalled) depends jointly on the solvent driving force $f_s$, the capillary diameter $d$, and the Young’s modulus $Y$, a full exploration of the three-dimensional phase space using explicit-solvent simulations is computationally expensive. To address this challenge, we first employ implicit-solvent simulations to efficiently map the translocation phase boundary in the $(d,f_s,Y)$ parameter space and to estimate critical values such as the minimum constriction diameter $d_c$ and driving force $f_c$ required for successful entry. These results are then complemented by explicit-solvent simulations, which capture hydrodynamic effects and are used to construct a detailed phase diagram for gels of fixed shear modulus ($G=0.03\,\epsilon/\sigma^3$) as functions of the solvent force $f_s$ and the constriction diameter $d$.

In equilibrium, \emph{i.e.} in the absence of external forcing, the microgel adopts a swollen configuration in the wide section of the channel (segment \textbf{a} in Fig.~\ref{fig:schematic_tube}). A driving force $f_s$ is then applied to the solvent molecules. In the absence of a gel or when the gel does not obstruct the channel, the solvent flows steadily through the corrugated geometry along the forcing direction. Owing to continuity and pressure--velocity coupling, the flow accelerates in regions of reduced cross section, accompanied by a pressure drop within the constriction. To characterize this baseline flow, we perform simulations of solvent transport through the empty channel under a representative driving force $f_s=0.1\,\epsilon/\sigma$. Particle velocities are expressed in cylindrical coordinates to exploit the axial symmetry of the geometry, and the axial velocity component is averaged over transverse directions. The channel is discretized in both axial and radial coordinates, and local velocity and virial stress are computed at each grid point to reconstruct the velocity and pressure fields (SI Fig.~4). As expected, the flow velocity is maximal within the constricted region [SI Fig.~4(a)], while the pressure attains a minimum there [SI Fig.~4(b)]. These qualitative features persist over a broad range of driving forces and channel geometries.

The presence of a microgel substantially modifies this flow field. For sufficiently strong driving, the gel deforms and translocates, allowing solvent to pass through the constriction. In contrast, below a critical driving force $f_s$, the gel stalls at the constricted region, leading to clogging and a pronounced reduction of solvent flux through the channel.

We begin by applying a weak driving force $f_s = 10^{-3}\epsilon/\sigma$ to each solvent particle along the tube axis, such that the microgel is advected with the solvent flow. Upon reaching the entrance of the constricted segment, the gel stalls and its center-of-mass velocity vanishes. If the gel remains immobile after a waiting time of $10^5\,\tau_{\mathrm{LJ}}$, the solvent force is incremented by $\Delta f_s = 10^{-3}$. This procedure is repeated until the microgel overcomes the entry barrier and enters the constricted channel. As shown in SI Fig.~10, illustrates this protocol for a microgel with shear modulus $G \approx 0.003\,\epsilon/\sigma^3$, with constriction diameters ranging from $17\sigma$ to $25\sigma$.
\begin{figure}[!h]
\centering
\includegraphics[width=\linewidth]{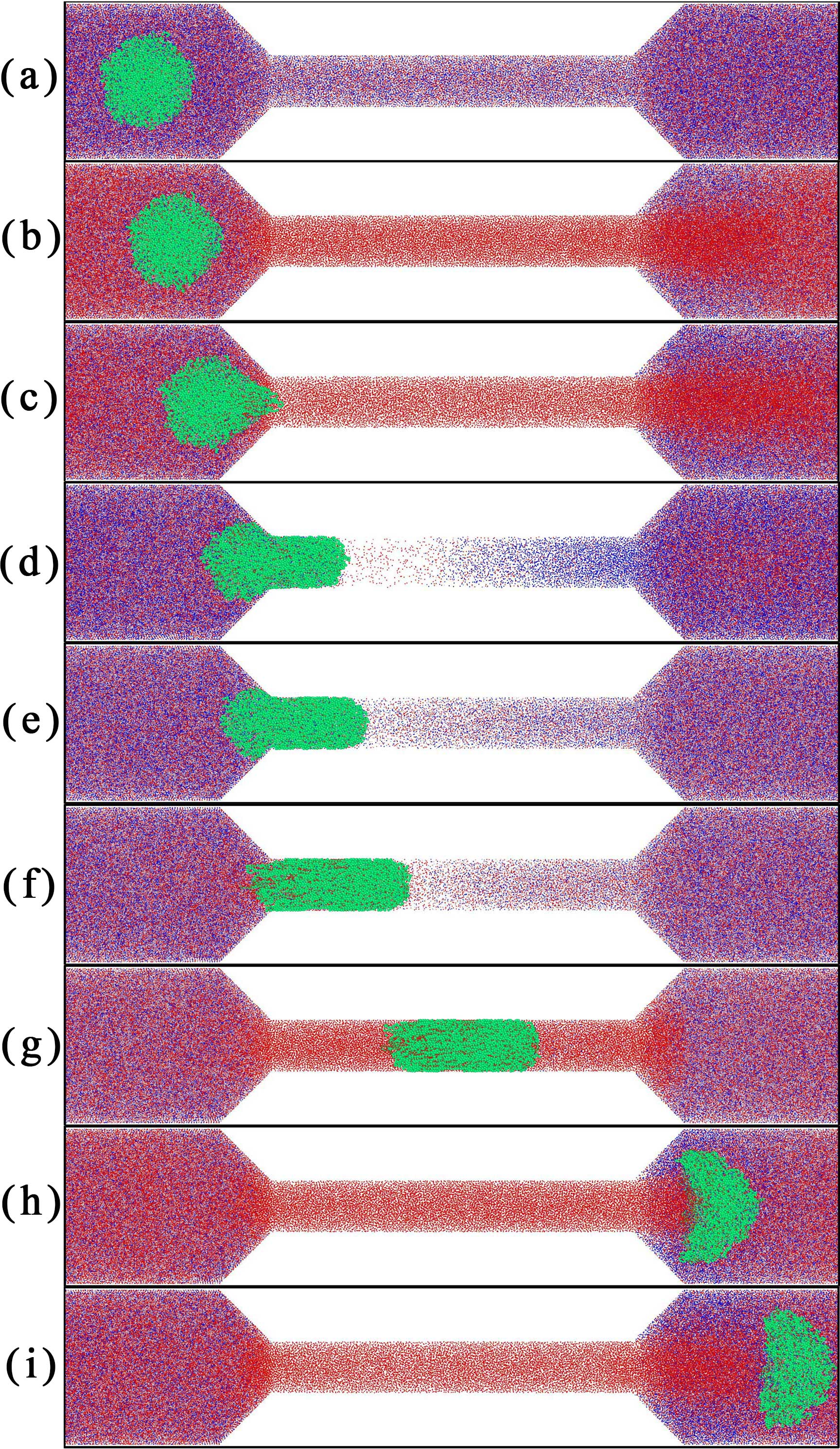}
\caption{Translocation of a crosslinked polymer microgel (green beads) through a corrugated channel in a good solvent with explicit solvent particles. The channel consists of two cylindrical segments of diameters $D = 60\sigma$ and $d = 20\sigma$ connected by a tapered segment. A constant force $f_s = 0.05\,k_B T/\sigma$ is applied to each solvent particle along the channel axis (left to right). The color scale from blue to red denotes increasing solvent velocity. Panel (a) shows the unperturbed microgel prior to forcing, while panels (b)--(i) illustrate successive stages of gel translocation through the corrugated channel under solvent-driven flow.}
\label{fig:successful_gel_transport}%
\end{figure}
To quantify gel transport, we track the axial position of the front meniscus, $x_f$, as a function of the applied solvent force $f_s$ ( SI Fig.~10). The minimum imposed force is $f_s=0.01\,\epsilon/\sigma$. Gels readily translocate through sufficiently wide constrictions, while narrower constrictions require larger driving. Owing to the imposed periodic boundary conditions, the temporal evolution of $x_f$ is periodic: following a successful translocation, the gel re-enters the channel from the upstream side, resulting in a discontinuous jump of $x_f$ from $300$ to $0$ in SI Fig.~10. Repeated translocation events enable reliable statistics of translocation times.

\begin{figure*}
\centering
\includegraphics[width=\linewidth]{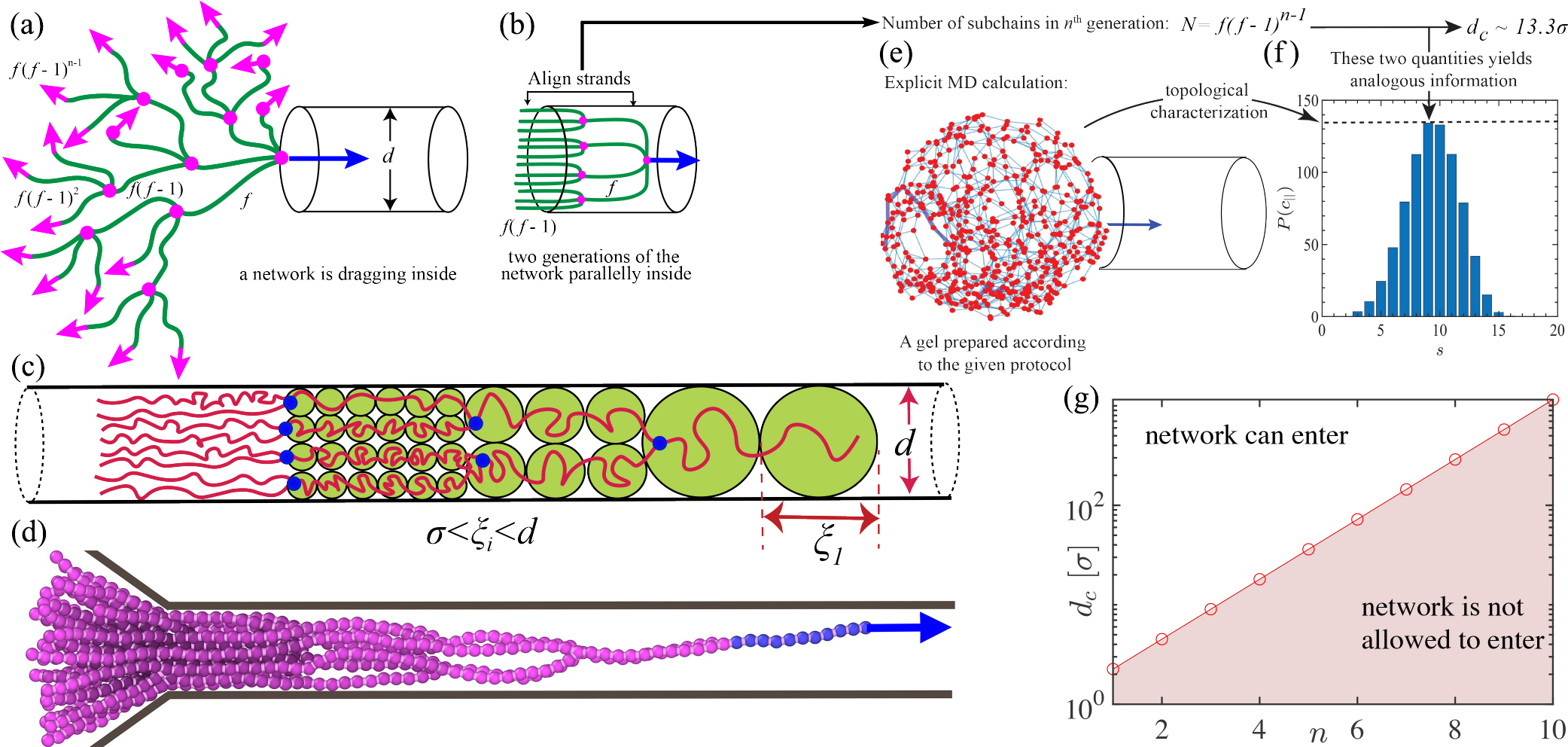}
\caption{
Schematic illustration of an ideal flexible Bethe branched chain entering a cylindrical tube of diameter $d$. 
(a) Network strands (green) connected by fixed nodes (magenta) of functionality $f$. 
(b) Under confinement, strands reorient and align parallel to the tube axis. 
(c) Blob representation of a dendron chain inside a narrow capillary, where the blob size $\zeta_i$ decreases with generation index while satisfying $\sigma < \zeta_i < d$. 
(d) Simulation snapshot of an $n=6$ dendron stalled at the tube entrance, consistent with the cutoff condition derived in SI Eq.~15. 
(e) Schematic of the topological characterization of a numerically synthesized gel network. The peak of the parallel connectivity distribution provides an effective measure of the number of load-bearing subchains in the $n^{\mathrm{th}}$ generation, which is used to estimate the critical translocation diameter shown in (f). 
(g) Phase diagram showing the maximum generation of an infinite dendron that can enter a tube of diameter $d_c$, as determined from SI Eq.~15. The boundary follows $\ln d_c = n_0 + \alpha n$.
}
\label{fig:bethelattice_in_tube}
\end{figure*}

Successful entry requires the gel to deform and squeeze into the constriction. In this state, the gel is compressed in the transverse directions, with polymer strands preferentially aligned along the tube axis, allowing efficient transport through the narrow region under minimal additional driving. Representative snapshots of a successful translocation are shown in Fig.~\ref{fig:successful_gel_transport}, where a longitudinal cross section of the channel is displayed. Gel beads are shown in green, while solvent particles are colored according to their axial velocity, from blue (low) to red (high).
When the gel does not obstruct the channel, the solvent velocity is enhanced within the constricted region relative to the wider sections. In contrast, when the gel stalls and clogs the constriction, solvent flow is strongly suppressed throughout the channel. Once the gel is fully inside or completely beyond the constricted segment, solvent flow is restored, producing a pronounced jet at the constriction exit. This jet deforms the trailing interface of the gel, leading to a discoid rear meniscus whose shape depends on the solvent velocity and gel stiffness.

\section{Results and Discussions}

Confinement of polymers within cylindrical geometries provides fundamental insight, within the blob picture framework, into the behavior of linear, branched, and star-like macromolecules under strong spatial restriction. Under such confinement, chain conformations result from a competition between axial stretching imposed by the tube and excluded-volume interactions between monomers. This balance governs the effective longitudinal extension of the chain as well as the spatial arrangement of strands within the tube cross-section. 

By imposing a close-packing constraint on $n_{||}$ parallel strands within the cylindrical cross-section, one obtains a geometric lower bound for the admissible tube diameter. In dimensionless units (see SI), this yields the critical cutoff diameter $d_c = 2 \sqrt{n_{||}/\pi}$,
which represents the minimal confinement size capable of accommodating $n_{||}$ parallel load-bearing strands without overlap. This cutoff therefore establishes a geometric threshold for compressing crosslinked networks into narrow capillaries.

\begin{figure*}[t!]
\centering
\includegraphics[width=\linewidth]{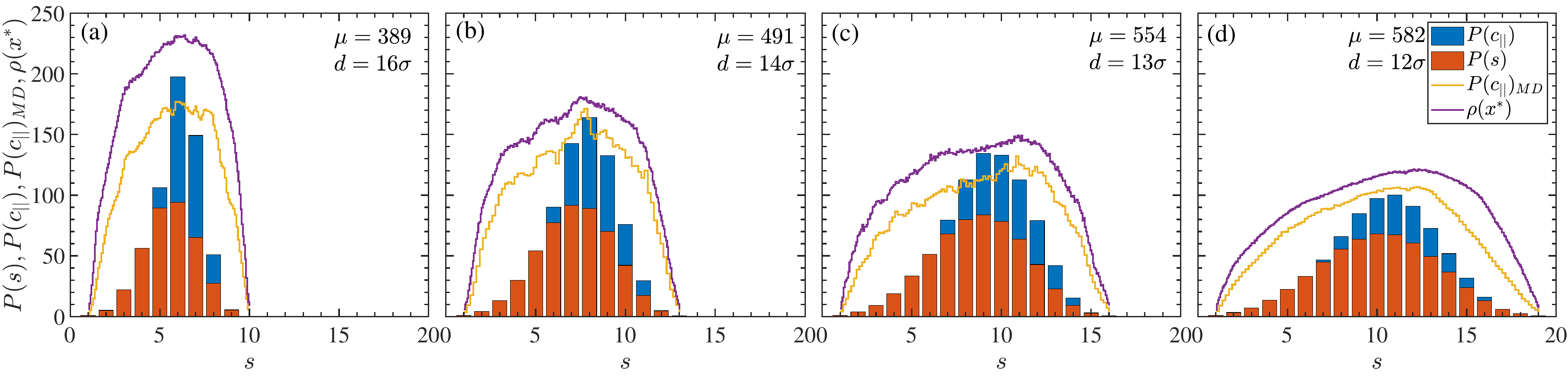}
\caption{
Mean shortest-path distribution of the gel networks, $P(s)$, and the mean parallel connectivity distribution, $P(c_{||})$, resolved at each generation as a function of path length for different gels (a–d). 
The monomer density distribution along the tube axis, $\rho(x)$, and the number of load-bearing subchains per axial bin obtained from MD simulations, $P(c_{||})_{\mathrm{MD}}$, are measured while the gel resides inside a narrow capillary with radius slightly above the critical cutoff diameter. 
A direct correspondence is observed between the intrinsic parallel connectivity $P(c_{||})$ and the axial subchain count from simulations, $P(c_{||})_{\mathrm{MD}}$, demonstrating that network topology dictates the effective strand organization under confinement. 
This agreement confirms that the gel is able to translocate through the constriction. 
The elasticity $Y$ and the number of network nodes $\mu$ are systematically varied from (a) to (d) to test the robustness of this correspondence across gels of differing mechanical and topological properties.
}%
\label{fig:shortest_path_p20_1_5_10_15}%
\end{figure*}
\subsection{Dendron Drag into a Narrow Capillary}

We next consider the translocation of a dendron (Bethe chain) whose core is dragged through a narrow cylindrical capillary by an externally applied force. As illustrated in Fig.~\ref{fig:bethelattice_in_tube}, an infinite-generation Bethe chain is driven into the tube by pulling its central backbone.

The previous geometric argument can be generalized to an ideal Bethe lattice of functionality $f$. At the $n^{\mathrm{th}}$ successor generation, the total number of strands is $n_{||} = f (f-1)^{n-1},$
as shown schematically in Fig.~\ref{fig:bethelattice_in_tube}(a). Under strong confinement, these strands are assumed to align approximately parallel to the tube axis (Fig.~\ref{fig:bethelattice_in_tube}(b)). 

The critical tube diameter is obtained by imposing a close-packing constraint on these $n_{||}$ parallel strands within the cylindrical cross-section. Equating the cross-sectional area available in the tube to the area occupied by the packed strands yields the cutoff condition for successful entry. Thus, for a cylinder of length $N\sigma$, the dendron can enter only if the diameter satisfies the critical condition derived previously.

Figure~\ref{fig:bethelattice_in_tube}(d) delineates the region in parameter space where the Bethe lattice cannot enter a tube of diameter below the critical threshold. The above geometric argument can be extended to include excluded-volume effects, providing a refined estimate of the critical radius beyond which the network cannot be compressed into the capillary. In this analysis, the solvent volume is neglected.

To validate this prediction, we perform simulations of a dendrimer with functionality $f=3$ and generation number $n$, which is pulled into a narrow capillary. For each fixed generation, the constriction diameter $d_c$ is varied systematically. Cross symbols indicate stalled configurations, whereas circular symbols denote successful entry of the $n^{\mathrm{th}}$ generation dendrimer into the channel.

\subsection{Phase Diagram of Successful Microgel Transport}

To extend the geometric cutoff framework to realistic network architectures, it is essential to quantify the topological connectivity between network junctions. We first analyze the shortest-path distribution between all node pairs in a given gel network. The resulting distribution $P(s)$ exhibits an approximately Gaussian form, whose width and median increase with the cavity volume in which the gel is synthesized (see Fig.~\ref{fig:shortest_path_p20_1_5_10_15}). Networks synthesized in larger cavities therefore display broader path-length distributions and correspondingly lower elastic moduli, reflecting softer and more weakly connected structures. While the shortest-path distribution captures the minimal topological distance between nodes, it does not fully describe the hierarchical organization relevant for confinement-driven transport (see Fig.~\ref{fig:bethelattice_in_tube}). In an idealized dendritic network, the number of shortest paths at a given generation defines a hierarchical level, with the associated histogram representing the number of strands in that generation. Under strong confinement, these strands would align and form a bundled structure within the constriction (Fig.~\ref{fig:bethelattice_in_tube}b). 

In contrast, real randomly crosslinked networks deviate from this idealized picture. Although subchains tend to align within the constriction, crosslinks between different generations introduce additional connectivity pathways (see SI Fig.~1). Consequently, the number of parallel load-bearing subchains cannot be inferred solely from shortest-path statistics. 
To address this, we develop a numerical procedure to compute the distribution of parallel connectivity, $P(c_{||})$, for a given network. Unlike the shortest-path distribution, which measures the minimal number of steps required to connect two nodes, the parallel connectivity distribution quantifies the number of distinct subchains contributing to load transmission at each hierarchical level or path index. 
In Fig.~\ref{fig:shortest_path_p20_1_5_10_15}, we compare the shortest-path distribution $P(s)$ (where $s$ denotes the path index) with the parallel connectivity distribution $P(c_{||})$. To validate the relevance of this metric under confinement, we analyze MD simulations of gels that successfully enter a capillary with diameter slightly above the critical threshold $d_c$. From these simulations, we compute the number of parallel subchains along the tube axis, $P(c_{||})_{\mathrm{MD}}$, by discretizing the tube into axial bins and counting load-bearing strands within each bin.
The monomer density distribution $\rho(x)$ is obtained by counting monomers within each axial bin. For comparison across different network architectures, we normalize the axial coordinate by the maximum generation number of the network, defining a dimensionless coordinate $x^*$. Thus, $\rho(x)$ is mapped to $\rho(x^*)$, enabling direct comparison with the generation index $s$ and connectivity distributions.
Here, $x$ has dimensions of length, whereas $x^*$ is dimensionless and analogous to the generation index $s$. The maximum axial extension of the gel inside the tube is normalized by the maximum number of generations connecting the two extremities of the translocating network.

As shown in Figs.~\ref{fig:shortest_path_p20_1_5_10_15}(a–d), the peak of the parallel connectivity distribution $P(c_{||})$ closely matches the peak of the axial subchain distribution obtained from simulations, $P(c_{||})_{\mathrm{MD}}$, when the gel resides inside a tube with diameter slightly above the critical threshold $d_c$. In this regime, the network can deform and reorganize such that its effective parallel strand count satisfies the geometric constraint for entry.
For larger capillary diameters, the required deformation decreases and additional strands can enter the constriction simultaneously. Consequently, the waiting time of the gel at the tube entrance is reduced.
In this analysis, only parallel load-bearing connections are considered; series connections between nodes at a given generation do not contribute to the effective transverse packing constraint. The monomer density distribution $\rho(x^*)$ along the tube axis is also shown in Figs.~\ref{fig:shortest_path_p20_1_5_10_15}(a–d) for tubes with diameter slightly above $d_c$. For tightly interconnected networks with abundant series connections (e.g., Fig.~\ref{fig:shortest_path_p20_1_5_10_15}(a)), the peak position of $P(c_{||})_{\mathrm{MD}}$ deviates from that of $\rho(x^*)$, reflecting additional internal connectivity that does not directly contribute to transverse strand packing.
The peak value of $P(c_{||})$ provides an effective estimate of the number of parallel load-bearing strands, $n_{||}$, and is therefore used in place of the idealized strand count in SI Eq.~15 to obtain an improved estimate of the critical diameter $d_c$. Alternatively, the maximum monomer count per axial bin from MD simulations can be used to estimate $d_c$ via SI Eq.~15.

Thus, the ability of a given gel to enter a narrow capillary can be predicted directly from its network architecture. The critical tube radius obtained from the connectivity model agrees closely with the diameter at which the gel marginally enters the constriction in simulations. This procedure enables identification of the geometric phase boundary separating the mobile regime from the mechanically arrested state.
However, geometry alone does not fully determine translocation. The applied pressure (or force) plays a decisive role in overcoming elastic resistance during entry. If the equilibrium gel diameter $2R_{\mathrm{gel}}$ is smaller than or comparable to the constriction diameter $d$, the gel enters without significant additional pressure. For $d > d_c$, gels with equilibrium radius $R_{\mathrm{gel}} > d/2$ can still enter under sufficient applied force. Conversely, for $d < d_c$, entry is prohibited irrespective of the applied pressure due to topological constraints.

\begin{figure}[!t]
 \centering
 \includegraphics[width=\linewidth]{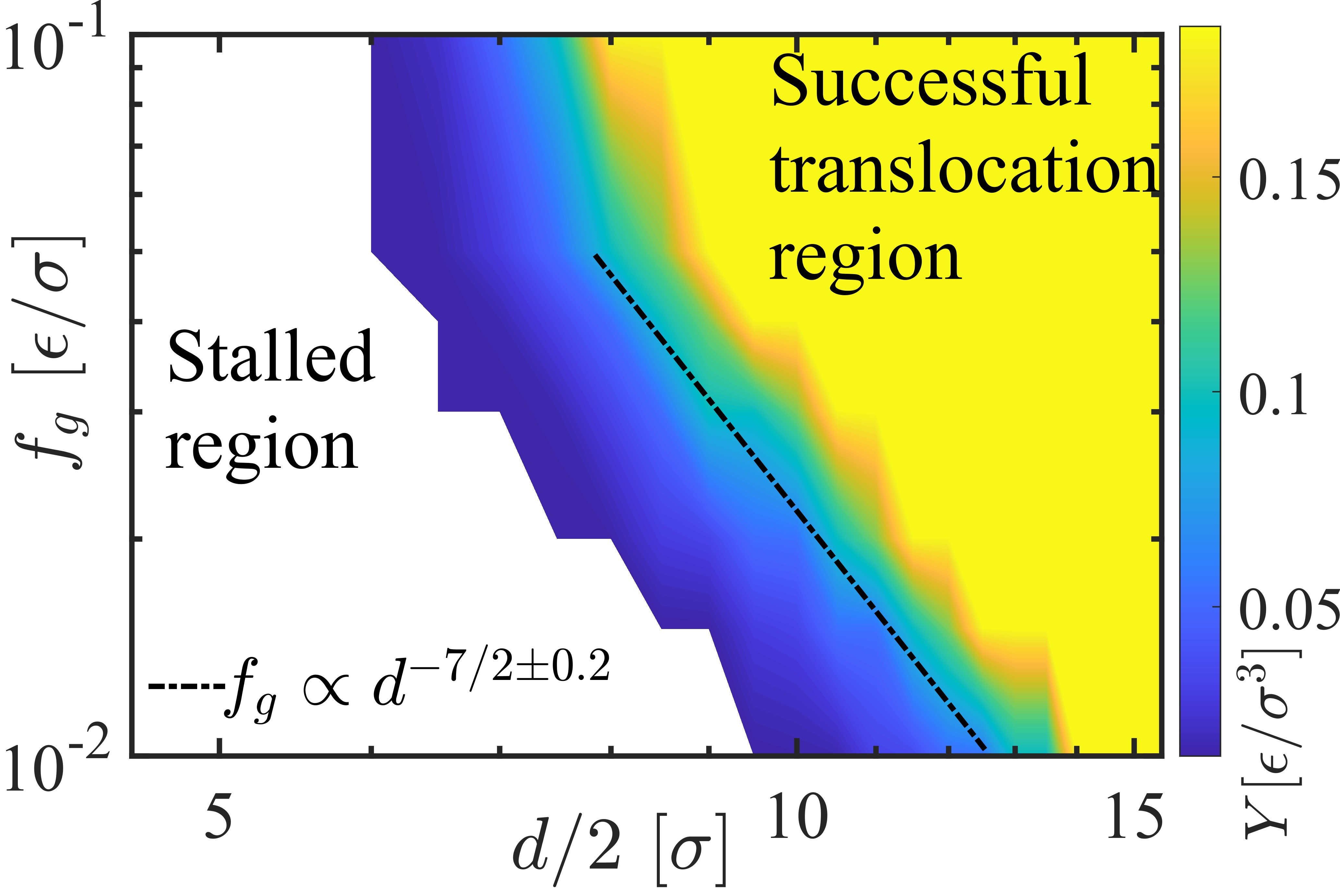}
 \caption{ Phase diagram of the successful gel translocation for different radii $d/2$ of the constricted capillaries with force on gel $f_g$ for different Young's moduli, $Y$, of gels.
Data are shown on a logarithmic scale to highlight scaling behavior (see SI Fig. 12 for the 3D phase diagram). 
The solid boundary separates the mobile and mechanically arrested regimes and follows the power-law relation 
$f_g \propto d^{-7/2 \pm 0.2}$ over the accessible range of parameters.}
\label{fig:transport_3d_phase_diagram_log}
\end{figure}

To quantify the combined effects of geometry and forcing, we perform CGMD simulations of microgels driven by an external force $f_g$ inside the corrugated tube shown in Fig.~\ref{fig:schematic_tube}. Solvent-mediated transport would require substantially larger system sizes and computational cost; therefore, we focus on force-driven translocation to determine the conditions for successful passage. We construct phase diagrams in the $(f_g, d)$ plane for gels with different Young’s moduli $Y$, as shown in Fig.~\ref{fig:transport_3d_phase_diagram_log}. The constriction diameter is varied in the range $10\sigma \le d \le 30\sigma$, while geometric parameters $L_1$ and $L_2$ are adjusted to maintain a fixed opening angle.

We systematically vary the constriction diameter $d$ and the applied force $f_g$ to construct a two-dimensional phase diagram of successful translocation for each gel architecture. CGMD simulations are performed for gels with different Young’s moduli $Y$ to determine how elastic stiffness shifts the phase boundary (Fig.~\ref{fig:transport_3d_phase_diagram_log}). 
In SI Fig.~12a, red crosses denote parameter combinations for which the gel fails to enter the constriction, while colored circles indicate successful translocation. These two regimes are separated by a continuous surface boundary in the $(d, f_g, Y)$ parameter space. A contour representation of this three-dimensional phase diagram is shown in SI Fig.~12b, where color encodes the Young’s modulus in the range $0.02 \le Y \le 0.18\,\epsilon/\sigma^3$.
As expected, stiffer gels (larger $Y$) require higher applied forces to translocate through a given constriction, whereas softer gels can enter narrower tubes under comparatively smaller forcing. Increasing $f_g$ enables a given gel to overcome elastic resistance and enter progressively smaller constrictions, provided $d > d_c$.

When plotted on logarithmic axes (Fig.~\ref{fig:transport_3d_phase_diagram_log}), the phase boundary exhibits a robust power-law scaling, $f_g \propto d^{-7/2 \pm 0.2}$, across a broad range of gel stiffnesses. This scaling highlights the coupled role of geometry and elasticity in determining the onset of translocation.

\begin{figure}[!t]
\centering
\includegraphics[width=\linewidth]{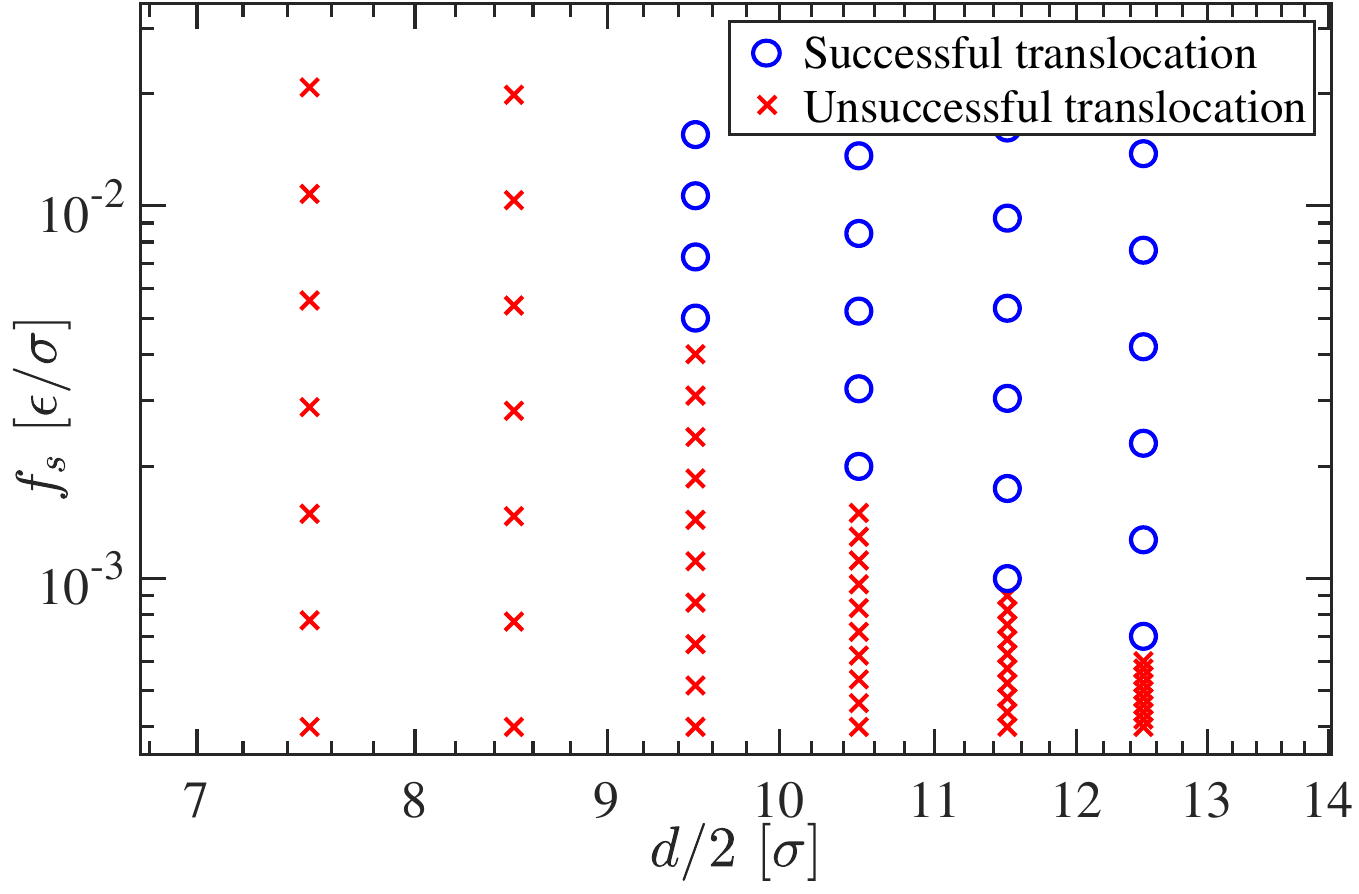}
\caption{Solvent-mediated phase diagram of microgel translocation for a gel with fixed shear modulus $G = 0.3\epsilon/\sigma^3$. 
The applied force on solvent particles, $f_s$, is plotted against the constriction diameter $d$. 
Blue circles denote successful translocation, while red crosses indicate mechanically arrested states. 
The solid boundary delineates the transition between mobile and stalled regimes.}
\label{fig:solvent_mediated_phase_diagram_01}%
\end{figure}
To complement the force-driven mechanism, we also investigate solvent-mediated transport. Due to computational constraints, these simulations are restricted to a single gel modulus ($G = 0.3\epsilon/\sigma^3$), as shown in Fig.~\ref{fig:solvent_mediated_phase_diagram_01}. Instead of directly pulling the gel, a force $f_s$ is applied to solvent particles. Blue circles denote successful entry, whereas red crosses indicate stalled configurations.
We find that even for sufficiently large solvent forcing ($f_s > 0.1\,\epsilon/\sigma$), the gel cannot enter constrictions with diameter $d < 18\sigma$. The connectivity model predicts a critical diameter $d_c \approx 17\sigma$, in close agreement with simulation results. The transition boundary in the solvent-driven case follows approximately
$f_s \sim d^{-4 \pm 0.5},$
which slightly deviates from the implicit force-driven scaling, reflecting differences in stress transmission mechanisms.

\subsection{Volume and Shape Change During Translocation}

In a typical translocation simulation, a small force is first applied to the solvent particles and subsequently increased incrementally. For each force value $f_s$, the system is allowed to relax until the position of the gel tip, $x_f$, reaches a steady state. Below the critical force, the microgel adopts a stationary equilibrium configuration at the entrance of the constriction. When $f_s$ exceeds the threshold value, the gel exhibits a finite loading time before entering the narrow capillary. This loading time depends on the applied force, constriction diameter, and gel elasticity.
\begin{figure}[!b]
    \centering \includegraphics[width=\linewidth]{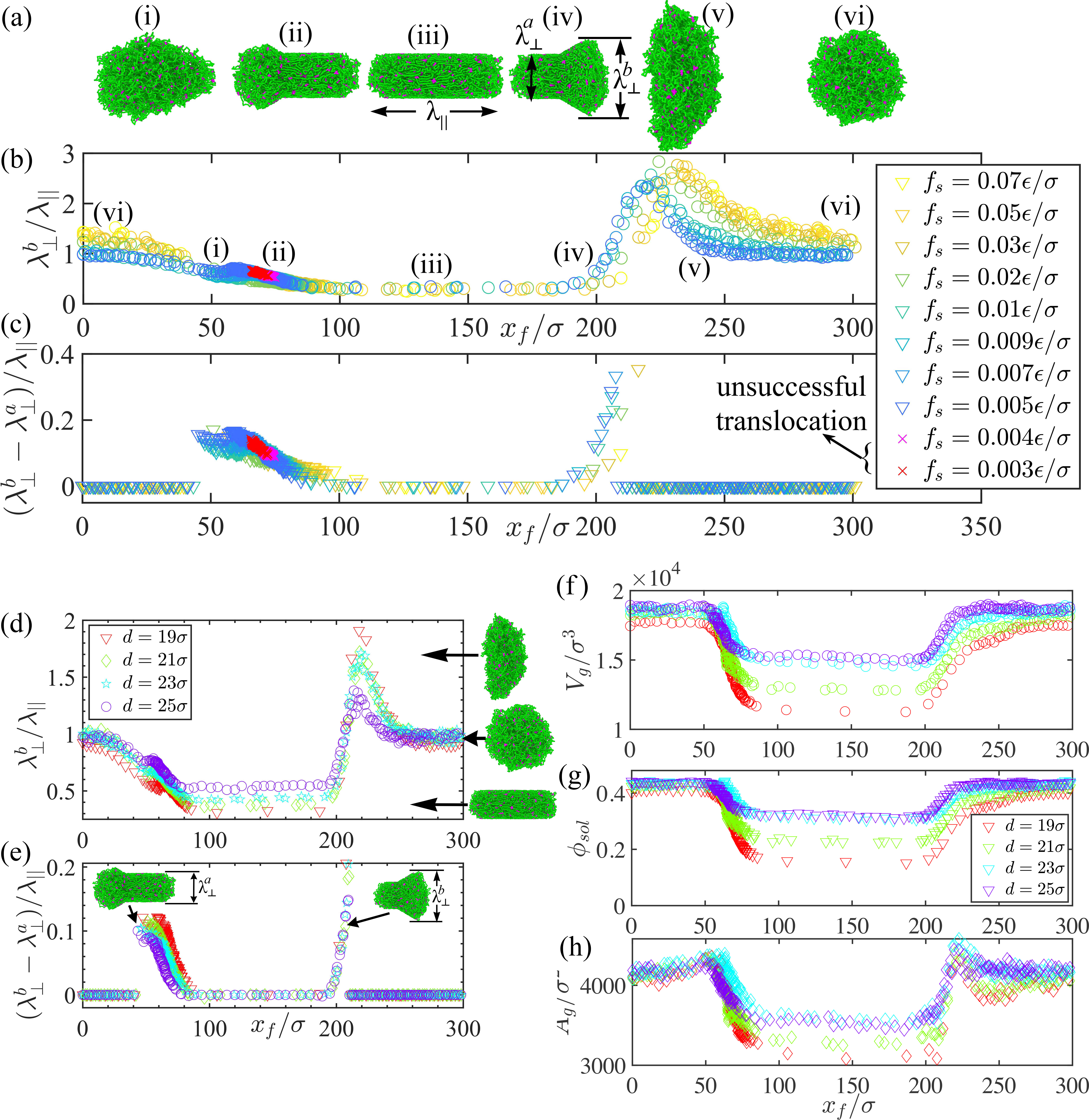} 
    \caption{(a) Simulation snapshots illustrating successive shape transformations of a microgel translocating through a corrugated tube with constriction radius $d_c/2 = 9.5\sigma$. 
    The longitudinal extent along the tube axis is denoted $\lambda_{||}$, while $\lambda_{\perp}^a$ and $\lambda_{\perp}^b$ represent orthogonal radial extents inside and outside the constriction, respectively. 
    (b) Deformation index $\lambda_{\perp}^b/\lambda_{||}$ as a function of tip position $x_f$ for different applied forces $f_s$ (dark to light denotes increasing force). Red crosses indicate stalled states. 
    (c) Asymmetry measure $(\lambda_{\perp}^b - \lambda_{\perp}^a)/\lambda_{||}$ during entry and exit. 
    (d–f) Evolution of gel volume $V_g$, solvent volume fraction $\phi_{\mathrm{sol}}$, and surface area during translocation.    }\label{fig:transport_gel_shape}
\end{figure} 
An initially spherical microgel placed in the solvent deforms under advection into a pear-like shape in the funnel region (Fig.~\ref{fig:transport_gel_shape} a, i). As the gel is progressively squeezed into the constriction, it undergoes a continuous transformation from a bouquet-like configuration (Fig.~\ref{fig:transport_gel_shape} a, ii) to an elongated cylindrical shape (Fig.~\ref{fig:transport_gel_shape} a,iii). Once fully confined, the gel maintains a cylindrical morphology throughout its passage inside the narrow capillary.

Upon exiting the constriction, elastic relaxation and solvent influx cause the gel to re-expand. The precise exit morphology depends on the interplay between fluid velocity and gel elasticity. If the kinetic energy of the emerging solvent jet exceeds the elastic resistance of the gel, a transient discoid deformation is observed; otherwise, the gel gradually recovers its spherical shape in the wider downstream section (Fig.~\ref{fig:transport_gel_shape}a,vi).

To quantify shape evolution, we define the longitudinal extent along the tube axis as $\lambda_{||}$ and two orthogonal radial extents as $\lambda_{\perp}^a$ (inside the constriction) and $\lambda_{\perp}^b$ (outside the constriction). When the gel is fully confined or fully outside the constriction, $\lambda_{\perp}^a = \lambda_{\perp}^b$. During partial entry or exit, however, $\lambda_{\perp}^b > \lambda_{\perp}^a$, reflecting asymmetric deformation.

We define a deformation index as
$\frac{\lambda_{\perp}^b}{\lambda_{||}},$
which captures the aspect ratio of the gel during transport. Figure~\ref{fig:transport_gel_shape}b shows the evolution of this index as a function of tip position $x_f$ for different applied forces $f_s$ (dark to light colors denote increasing force). For successful translocation, $\lambda_{\perp}^b/\lambda_{||} < 1$ while the gel is fully confined. Unsuccessful transport at low $f_s$ is indicated by red crosses.
To further resolve entry and exit asymmetry, we compute $(\lambda_{\perp}^b - \lambda_{\perp}^a)/\lambda_{||},$ which vanishes when the gel is entirely inside or outside the constriction and becomes non-zero during loading and release. Notably, the deformation amplitude is larger during exit than during entry, whereas the axial extent of the loading region exceeds that of the release region.
The deformation index depends strongly on the constriction diameter $d$, as shown in Fig.~\ref{fig:transport_gel_shape}. 
Inside the narrow tube, $\lambda_{\perp}^b/\lambda_{||}$ increases with increasing $d$, since weaker confinement reduces axial stretching and allows the gel to retain a more isotropic shape (Fig.~\ref{fig:transport_gel_shape}d). This monotonic trend is consistent with the deformation observed at the channel entrance. 
In contrast, the exit region exhibits non-monotonic behavior: gels confined in narrower tubes undergo stronger deformation and display a more pronounced radial expansion upon release, whereas gels confined in wider tubes relax more gradually. Despite this difference, the shape change during loading and release (Fig.~\ref{fig:transport_gel_shape}e) is largely independent of $d$.

\begin{figure}[h!] 
    \centering\includegraphics[width=\linewidth]{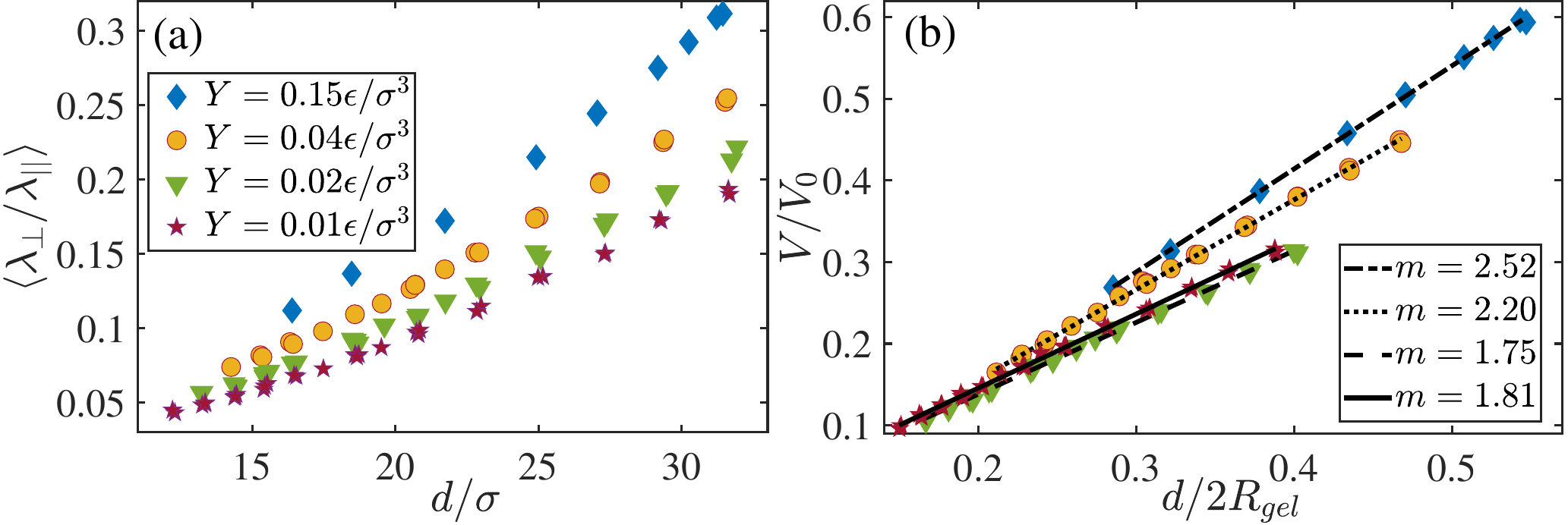} 
    \caption{
    (a) Deformation index $\langle \lambda_{\perp}/\lambda_{||} \rangle$ of fully confined gels as a function of constriction diameter $d$ for different Young’s moduli $Y$. 
    (b) Normalized gel volume $V/V_0$ as a function of the confinement parameter $d/2R_{\mathrm{gel}}$ for gels with varying $Y$. The approximately linear dependence indicates geometry-controlled compression in the explored regime.
    }
    \label{fig:relation_volume_change_with_tube_radius} 
\end{figure}

Confinement induces solvent expulsion from the gel network. The gel volume, solvent fraction, and surface area reach their minimum values when the microgel is fully confined inside the constriction (Fig.~\ref{fig:transport_gel_shape}f–h). We analyze a gel with equilibrium diameter $2R_{\mathrm{gel}} \approx 33\sigma$ translocating through constrictions of diameter $d=19\sigma$, $21\sigma$, $23\sigma$, and $25\sigma$. A monotonic decrease in gel volume is observed with decreasing $d$. The equilibrium solvent fraction is $\phi_{\mathrm{sol}} \approx 0.45$, which decreases upon entry due to solvent egress. The gel surface area similarly decreases under compression, followed by a transient increase during exit as the gel re-expands.
When the gel is fully confined inside the cylindrical constriction, we denote the axial extension as $\lambda_{||}$ and the radial extension as $\lambda_{\perp}$. The deformation index $\lambda_{\perp}/\lambda_{||}$ inside the constriction is shown in Fig.~\ref{fig:relation_volume_change_with_tube_radius}a. As confinement strengthens (smaller $d$), $\lambda_{||}$ increases due to biaxial compression, resulting in a decrease of $\lambda_{\perp}/\lambda_{||}$.

The normalized volume deformation ratio $V/V_0$ (where $V_0$ is the equilibrium gel volume) is plotted as a function of the dimensionless confinement parameter $d/2R_{\mathrm{gel}}$ in Fig.~\ref{fig:relation_volume_change_with_tube_radius}b for gels of different Young’s moduli $Y$. The data exhibit an approximately linear dependence of $V/V_0$ on $d/2R_{\mathrm{gel}}$ across the explored range. For strong confinement and sufficiently soft gels (e.g., $Y \sim 0.01\,\epsilon/\sigma^3$), the confined volume becomes nearly independent of $Y$, indicating a geometry-dominated regime.
The slope of $V/V_0$ versus $d/2R_{\mathrm{gel}}$ decreases with decreasing Young’s modulus, reflecting enhanced compressibility of softer gels. In contrast, Li {\it et al.}~\cite{p:li2015} reported a universal scaling behavior  $V/V_0 \propto (7/5)^{1/3} (d/2R_{\mathrm{gel}})^{4/3}$,independent of $Y$. The deviation observed here highlights the influence of network topology and finite extensibility in randomly crosslinked gels.

\begin{figure}[!h]
\centering
\includegraphics[width=\linewidth]{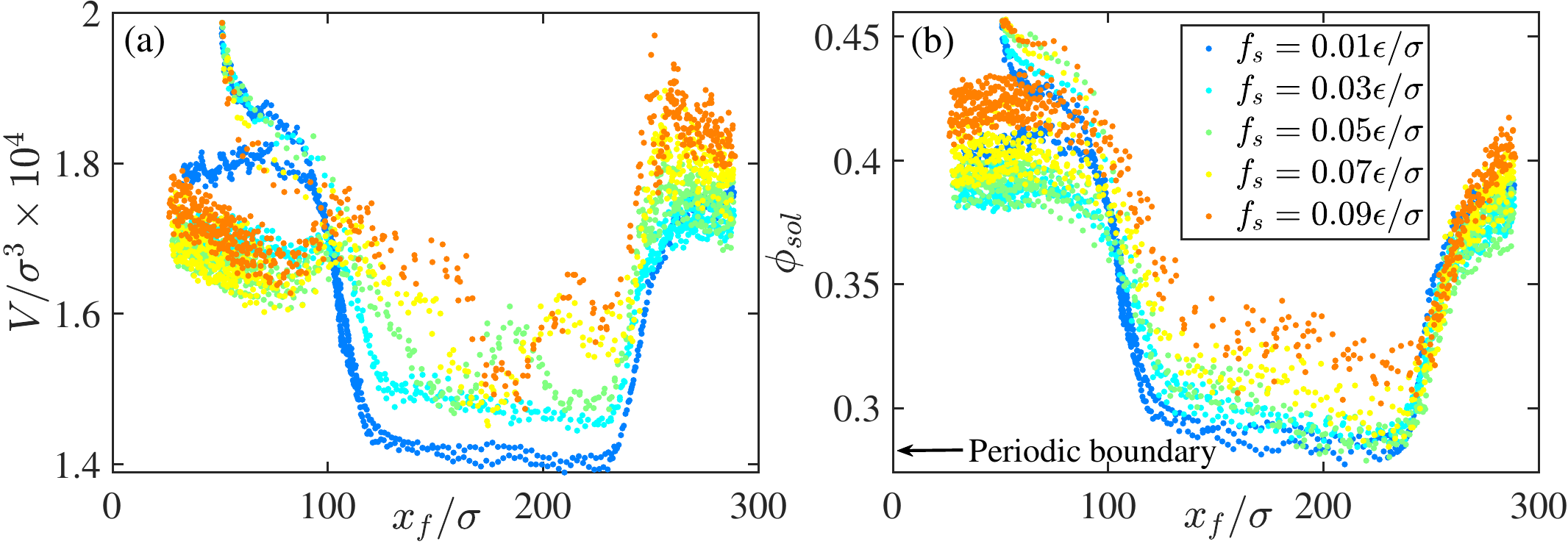}
\caption{
(a) Gel volume $V$ as a function of tip position during translocation through the corrugated tube. 
(b) Solvent volume fraction inside the gel, $\phi_{\mathrm{sol}}$, for different applied solvent forces $f_s$. 
Colors denote increasing $f_s$.
}
\label{fig:volume_of_gel}
\end{figure}

An additional feature of the translocation process is the force-dependent volume relaxation of the gel, shown in Fig.~\ref{fig:volume_of_gel}. At low applied solvent force $f_s$, the gel advances slowly and undergoes substantial solvent expulsion, resulting in a pronounced volume reduction. In contrast, higher forcing reduces the loading time and limits solvent egress, leading to a smaller overall volume change. Thus, slower transport allows greater structural relaxation and compression of the network.

We next quantify strand alignment within the constriction by computing the second Legendre polynomial order parameter,
\begin{equation}
\mathcal{P}_2 = \frac{1}{2}(3\cos^2\theta - 1),
\end{equation}
where $\theta$ is the angle between the end-to-end vector of a subchain and the tube axis. The order parameter satisfies $-1/2 \le \mathcal{P}_2 \le 1$, with $\mathcal{P}_2=1$ corresponding to perfect alignment along the axis and $\mathcal{P}_2=-1/2$ indicating orthogonal alignment (pancake-like deformation). Isotropic configurations yield $\langle \mathcal{P}_2 \rangle = 0$ \cite{p:nikoubashman2017, p:milchev2018}.
\begin{figure}[!h]
\centering
\includegraphics[width=\linewidth]{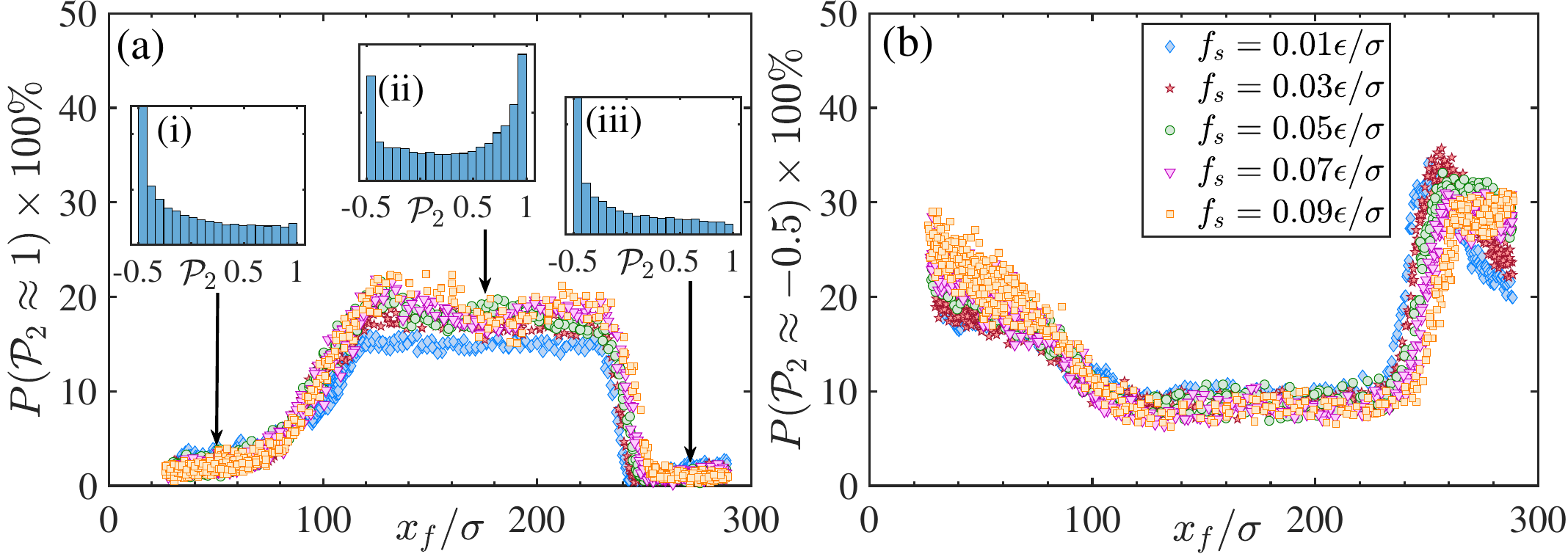}
\caption{
Orientational order of polymer strands inside the tube. 
(a) Distribution of $\mathcal{P}_2$ showing strong axial alignment ($\mathcal{P}_2 \approx 1$) under narrow confinement. 
(b) Distribution of $\mathcal{P}_2$ illustrating reduced orthogonal alignment ($\mathcal{P}_2 \approx -0.5$) inside the constriction.
}
\label{fig:nematic_like_order}
\end{figure}
To compute $\mathcal{P}_2$, the tube is discretized into axial bins of width $\sim 2\sigma$. For each bin, strands are identified and their normalized end-to-end vectors are evaluated relative to the tube axis. In wide tubes, strand orientations remain approximately isotropic, as shown in the insets of Fig.~\ref{fig:nematic_like_order}(a). Under strong confinement, however, the distribution develops a pronounced peak near $\mathcal{P}_2 \simeq 1$, indicating nematic-like alignment along the tube axis.
Conversely, the probability of orthogonal alignment ($\mathcal{P}_2 \simeq -0.5$) decreases sharply inside narrow constrictions (Fig.~\ref{fig:nematic_like_order}b), confirming that confinement suppresses transverse strand orientations and promotes axial ordering.

\subsection{Passage of Microgel Through a Constriction}

\begin{figure}[h!]
 \centering
 \includegraphics[width=0.9\linewidth]{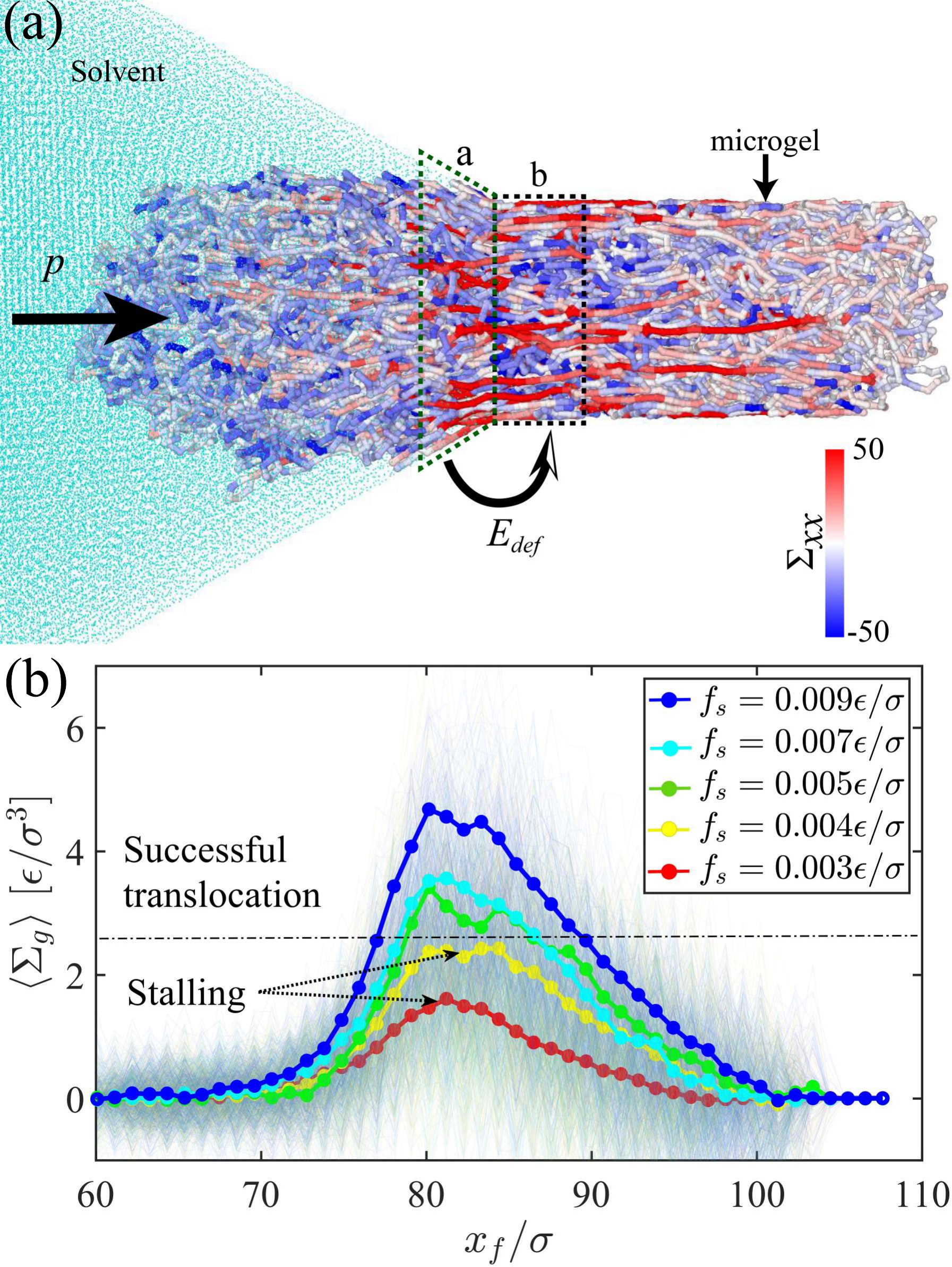}
 \caption{
 (a) Simulation snapshot of a microgel stalled at the entrance of a constriction. 
 Color indicates local strand stress (red: highly stressed; blue: weakly stressed). 
 Cyan particles represent solvent pushing the gel along the tube axis. 
 (b) Stress anisotropy $\langle \Sigma_g \rangle$ along the tube axis for different applied forces. 
 The peak occurs at the constriction entrance, corresponding to maximal elastic resistance.
 }
 \label{fig:clogged_gel}
\end{figure}

Successful translocation through a narrow capillary requires that the applied pressure exceeds the elastic resistance of the gel. In our simulations, the pressure difference is generated by applying a force $f_s$ to solvent particles. Translocation occurs only when $f_s > f_s^{c}$, where $f_s^{c}$ is the critical force required to drive a gel into a constriction of diameter $d$.
Figure~\ref{fig:clogged_gel}(a) illustrates a gel stalled at the entrance of the narrow capillary. The applied pressure compresses the upstream region (labeled (a)) and attempts to deform it into the constricted geometry (region (b)). This deformation requires an energy cost $E_{\mathrm{def}}$. When the applied pressure is insufficient, the gel remains arrested. Once the pressure exceeds the critical threshold, the gel overcomes this deformation barrier and enters the constriction. After passing the bottleneck, no additional excess pressure is required to sustain motion.

We observe three dynamical stages during translocation: 
(i) deformation of the initially spherical gel into a pear-like structure near the tapered orifice due to solvent advection; 
(ii) arrest of the rear portion, whose effective radius approaches $R_g \approx d/2$; 
(iii) full entry once the applied pressure exceeds the deformation threshold.
When stalled, the solvent flow rate decreases significantly \cite{p:li2020}. This behavior is consistent with the connectivity-based cutoff diameter derived earlier, below which entry is geometrically prohibited. However, for $d > d_c$, successful passage depends on the balance between applied pressure and internal gel stress.
Neglecting static wall friction and assuming slip boundary conditions \cite{p:wyss2010}, the longitudinal gel stress $\Sigma_g$ must balance the applied pressure $p$. The condition for translocation is therefore $ p > \Sigma_g $.
We compute the axial stress anisotropy as $
\Sigma_g = \Sigma_x - \frac{1}{2}(\Sigma_y + \Sigma_z),$
which peaks at the constriction entrance (Fig.~\ref{fig:clogged_gel}(b)). Increasing $f_s$ raises $\Sigma_g$, enabling the gel to elongate axially and enter the constriction.
Once fully confined, the gel is subject to axial normal stress $\Sigma_x$ and radial stress $\Sigma_r$. The disk experiences a net normal stress $\delta\Sigma_x$ and a shear traction $\tau_w$ along the axial direction~\cite{p:li2020}. Using a linear frictional constitutive relation,
\begin{eqnarray}
\tau_w =\tau_0 -c_1 \Sigma_r,
\end{eqnarray}
where $\tau_0$ depends on gel--wall adhesion and $c_1$ is a proportionality constant. Assuming a uniform hydrodynamic pressure acting on the upstream face, force balance gives
\begin{eqnarray}\label{eq:li_1}
\pi d^2 \mathrm{d}\Sigma_x = 2 \pi d \tau_w \mathrm{d}x, \\
\frac{\mathrm{d}\Sigma_x}{\mathrm{d}x} = \frac{2(\tau_0 - c_1 \Sigma_r)}{d},
\end{eqnarray}
where $\tau_w$ is the wall shear traction, $\tau_0$ is a gel–wall adhesion parameter, and $c_1$ is a proportionality constant.

For a neo-Hookean material under equibiaxial deformation,
\begin{eqnarray}\label{eq:li_2}
\Sigma_x - \Sigma_r = c_2 (\lambda_{||}^2 - \lambda_{\perp}^2),
\end{eqnarray}
where $c_2$ is related to the elastic modulus and $\lambda_{||}$ and $\lambda_{\perp}$ are the axial and radial stretch ratios.
Li {\it et al.}~\cite{p:li2020} solved these equations assuming incompressibility ($\lambda_{||} = \lambda_{\perp}^{-2}$). In contrast, our simulations show significant volume reduction under strong confinement (Fig.~\ref{fig:relation_volume_change_with_tube_radius}), indicating that the incompressible assumption is not strictly valid in this regime. Under extreme deformation, stress estimation based on this framework becomes unreliable. A more general constitutive model accounting for compressibility and finite extensibility is required and will be addressed in future work.

\subsection{Motion of the Microgel During Passage}

\begin{figure}[h!]
\centering
\includegraphics[width=\linewidth]{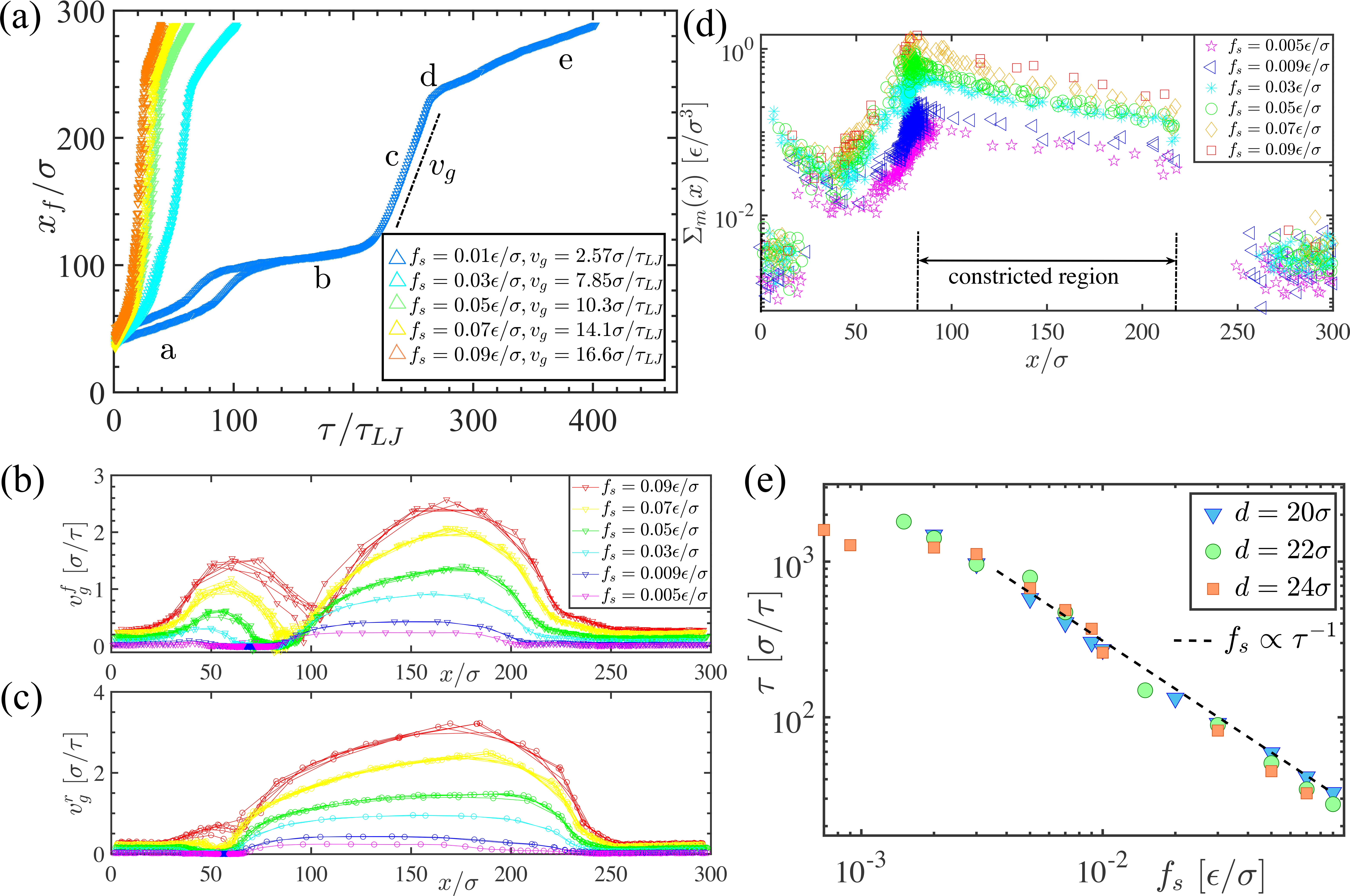}
\caption{
(a) Front meniscus position $x_f$ versus time $\tau$ for different applied solvent forces $f_s$. The slope in the constricted region yields the steady gel velocity. 
(b–c) Front and rear velocities ($v_g^f$ and $v_g^r$) as functions of applied force $f_s$. 
(d) Maximum stress anisotropy $\Sigma_m$ along the tube axis during translocation through a constriction of diameter $d=18\sigma$. Stress peaks at the entrance and decays exponentially inside the tube. 
(e) Translocation time $\tau$ versus applied force $f_s$ for different constriction diameters.
}
\label{fig:tip_distance_and_velocity}
\end{figure}

The microgel velocity $v_g$ determines the total translocation time through the constriction. Under slip boundary conditions and negligible wall friction, the velocity of the gel, once fully confined inside the narrow capillary, is governed primarily by the applied force $f_s$.
In solvent-mediated transport, we observe that the gel velocity inside the constricted region is approximately constant. Once fully encapsulated, the gel maintains a nearly steady conformation during transit (see Sec.~C), resulting in uniform translation along the tube axis.
Figure~\ref{fig:tip_distance_and_velocity}(a) shows the position of the front meniscus $x_f$ as a function of time $\tau$ for different applied solvent forces $f_s$. Three distinct velocity regimes are observed, corresponding to the wide entrance region, the tapered section, and the narrow constriction. The highest velocity occurs inside the constricted segment, whereas motion in the wider regions is slower, as evident from the slope of the displacement–time curves.

We separately track the front and rear menisci to quantify transient asymmetry. The velocities $v_g^f$ and $v_g^r$ are obtained from time derivatives of their respective positions. Near the constriction entrance, the front meniscus accelerates as it is drawn into the capillary, while the rear portion lags due to deformation and solvent resistance (Fig.~\ref{fig:tip_distance_and_velocity}b,c). This velocity mismatch reflects the progressive elongation of the gel during entry.

We also compute the maximum stress anisotropy $\Sigma_m$ along the tube axis. As shown in Fig.~\ref{fig:tip_distance_and_velocity}(d), $\Sigma_m$ peaks at the mouth of the constriction and decays approximately exponentially (slope $\approx -0.012$) inside the narrow segment. Outside the constricted and tapered regions, stress anisotropy vanishes as the gel relaxes toward an undeformed state. Increasing $f_s$ amplifies $\Sigma_m$, promoting axial elongation and facilitating entry.
Once fully confined, the gel traverses the constriction with nearly constant velocity. The total translocation time $\tau$ is shown in Fig.~\ref{fig:tip_distance_and_velocity}(e) for tube diameters $d=20\sigma$, $22\sigma$, and $24\sigma$. For a given force, $\tau$ is only weakly dependent on $d$, indicating that geometry primarily influences entry rather than steady-state motion. Translocation time decreases with increasing force, approximately following $\tau \propto \frac{1}{f_s}.$

The loading time $\tau_s$ defined as the time required for the gel to fully enter the constriction—depends sensitively on both $f_s$ and $d$. During loading, solvent egress and axial stretching transform the gel from spherical to cylindrical geometry. Note that $\tau_s$ decreases with increasing applied force (see SI Fig.~8), exhibiting a crossover from a non-monotonic regime at low forcing to an approximately linear scaling at higher forces.

\subsection{Stalled Microgel: Solvent Flow}

When a microgel becomes arrested at the entrance of a narrow capillary during solvent-mediated transport, the solvent flow behavior depends on the local gel volume fraction within the constriction. Two limiting scenarios may arise:

(i) High gel fraction $\phi_g$ at the bleb blocking the entrance, resulting from elevated solvent pressure insufficient to force complete entry. In this case, network pores collapse and fluid transport through the gel is suppressed.

(ii) Lower gel fraction $\phi_g$ under weaker forcing, where pores remain partially open and solvent can percolate through the stalled gel.

\begin{figure}[h!]
 \centering
 \includegraphics[width=\linewidth]{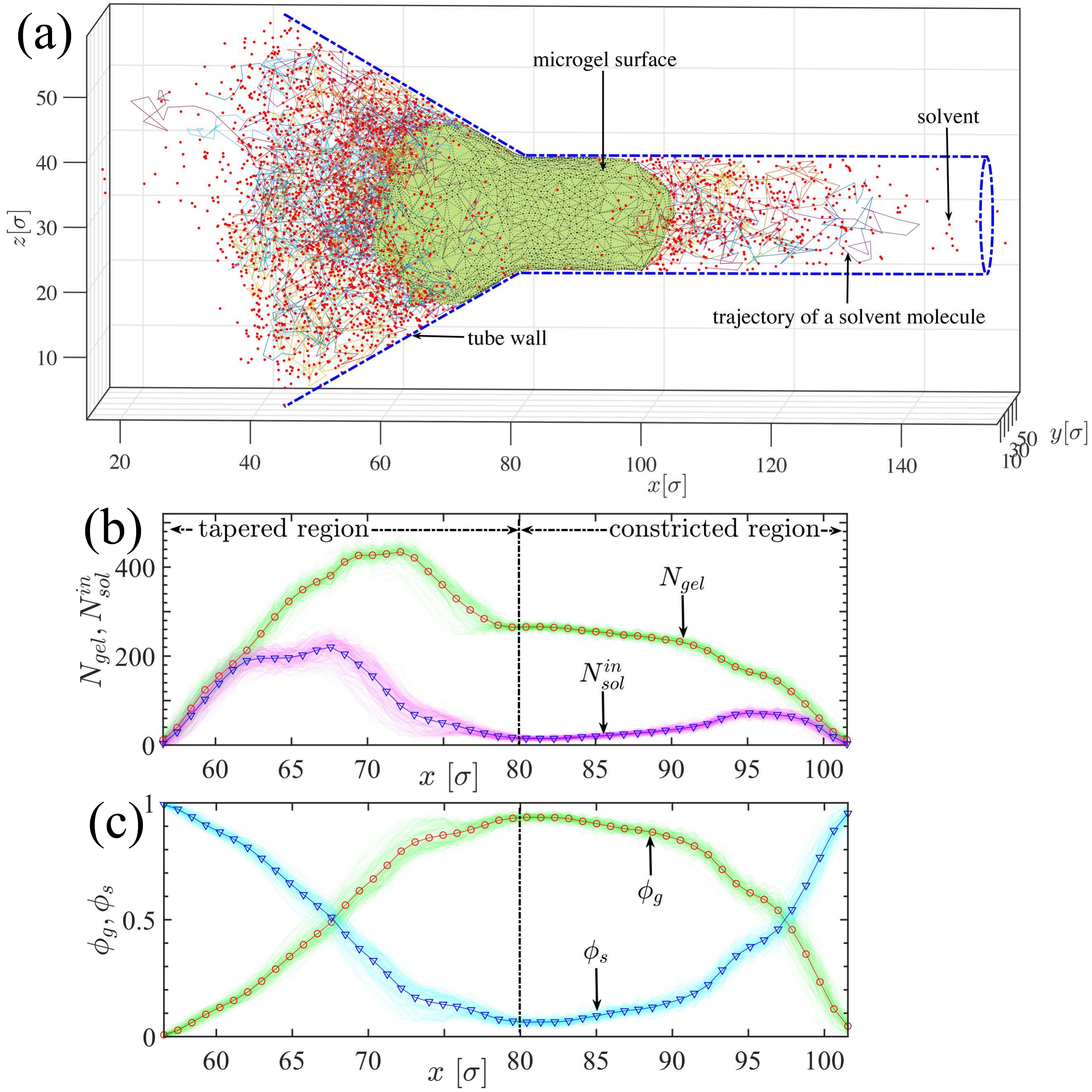}
 \caption{
 (a) Snapshot of a microgel clogging the constriction. Red dots represent solvent particles; colored lines show representative solvent trajectories. 
 (b) Axial distribution of gel monomers $N_{\mathrm{gel}}$ and solvent particles inside the gel, $N_{\mathrm{sol}}^{\mathrm{in}}$, for the stalled state. 
 (c) Gel and solvent volume fractions, $\phi_g$ and $\phi_s$, respectively. The vertical line marks the interface between the tapered and constricted regions. Transparent curves denote individual configurations; solid markers indicate ensemble averages.
 }
 \label{fig:stalling_region01}
\end{figure}

Figure~\ref{fig:stalling_region01}(a) shows a typical stalled configuration. Although the front meniscus has partially entered the constriction, the applied force ($f_s < f_s^c$) is insufficient for full translocation. In this state, no net axial solvent flux is observed ($Q \approx 0$). Instead, solvent particles exhibit random diffusive trajectories within the gel network.
To quantify solvent confinement, we identify gel boundaries using the $\alpha$-shape method \cite{p:stukowski2014} and compute the number of solvent particles enclosed within the gel, $N_{\mathrm{sol}}^{\mathrm{in}}$. The channel is discretized into axial bins to evaluate spatial distributions of gel monomers $N_{\mathrm{gel}}$, internal solvent particles, and total solvent density.
As shown in Fig.~\ref{fig:stalling_region01}(b,c), the gel fraction $\phi_g$ reaches a maximum at the constriction entrance, where solvent fraction $\phi_s$ is minimal. In this region, high local packing leads to effective pore closure and suppression of advective solvent transport. In contrast, the rear portion of the gel exhibits higher solvent content. Despite the absence of net flow, solvent molecules continue to diffuse within both front and rear sections of the arrested gel.

\section{Conclusion}

We have developed a unified framework to determine the conditions governing microgel entry into narrow capillaries and subsequent translocation through corrugated channels. By combining coarse-grained molecular dynamics simulations with scaling arguments based on excluded-volume interactions and network connectivity, we demonstrate the existence of a critical tube diameter $d_c$ below which a microgel cannot enter the constriction, irrespective of the applied force. Phase diagrams constructed in the $(f,d,Y)$ parameter space delineate regimes of successful translocation and mechanical arrest, and reveal that the qualitative structure of the phase boundary is robust across gels spanning a wide range of elastic moduli.
A central outcome of this work is the introduction of a connectivity-based topological criterion for gel passage. Using graph-theoretic analysis, we show that the maximum number of parallel subchains within the network controls the biaxial deformation required for cylindrical confinement. This leads to a direct relation between the critical capillary radius $d_c/2$ and the peak parallel connectivity, providing a predictive structural measure for confinement-induced arrest.

Hydrodynamically, the stalled state is characterized by vanishing axial solvent flux ($Q\approx 0$), consistent with pore closure at high local gel fractions. Upon successful entry, two transport regimes emerge: a weak-deformation regime with $Q \sim \Delta P$, and a strongly confined regime exhibiting nonlinear flow scaling~\cite{p:li2015,p:li2020}. These regimes reflect the interplay between elastic compression, solvent expulsion, and confinement geometry.
Confinement also induces pronounced structural reorganization. While the \textit{in silico} gels are isotropic in equilibrium, cylindrical confinement generates strong alignment of subchains along the tube axis, producing nematic-like order near $d_c$. During passage, the microgel undergoes systematic shape transitions quantified by axial extension $\lambda_{||}$ and radial extensions $\lambda_{\perp}^{a,b}$. The deformation index $(\lambda_{\perp}^a-\lambda_{\perp}^b)/\lambda_{||}$ provides a quantitative measure of shape asymmetry during loading and exit, and may serve as a mechanical signature of gel stiffness. Such metrics could be relevant for identifying mechanically abnormal soft objects, including pathological red blood cells, and for optimizing embolic gel design in biomedical applications~\cite{p:laurent2005}.

Stress analysis further reveals that axial stress peaks at the constriction entrance during arrest. Translocation occurs only when the applied pressure exceeds this stress barrier. Although our treatment captures the dominant axial response, a full tensorial characterization—particularly of radial stress distributions and non-affine deformations—remains an open challenge.
The present study establishes a direct link between microscopic network topology and macroscopic transport behavior under confinement. Future extensions incorporating explicit wall friction, adhesive interactions, and heterogeneous network architectures are expected to reveal additional dynamical regimes, including stick--slip motion and intermittent translocation. Developing continuum or network-level constitutive models capable of capturing the strongly compressible and non-affine deformations observed here will further advance the predictive theory of soft gel transport in biological and microfluidic environments.

\bibliographystyle{apsrev4-1}
\bibliography{bibliography}

@article{kramers1940brownian,
  title={Brownian motion in a field of force and the diffusion model of chemical reactions},
  author={Kramers, Hendrik Anthony},
  journal={physica},
  volume={7},
  number={4},
  pages={284--304},
  year={1940},
  doi={https://doi.org/10.1016/S0031-8914(40)90098-2},
  publisher={Elsevier}
}

@article{li2022fibrous,
  title={Fibrous hydrogels under biaxial confinement},
  author={Li, Yang and Li, Yunfeng and Prince, Elisabeth and Weitz, Jeffrey I and Panyukov, Sergey and Ramachandran, Arun and Rubinstein, Michael and Kumacheva, Eugenia},
  journal={Nature Communications},
  volume={13},
  number={1},
  pages={3264},
  year={2022},
  doi={https://doi.org/10.1038/s41467-022-30980-7},
  publisher={Nature Publishing Group UK London}
}

@article{p:hendrickson2010,
  author       = {Hendrickson, G. R. and Lyon, L. A.},
  title        = {Microgel Translocation through Pores under Confinement},
  journal      = {Angewandte Chemie International Edition},
  year         = {2010},
  volume       = {49},
  pages        = {2193--2197},
  doi          = {10.1002/anie.200906606}
}

@article{p:carugo2012,
  author       = {Carugo, D. and Capretto, L. and Willis, S. and Lewis, A. L. and Grey, D. and Hill, M. and Zhang, X.},
  title        = {A Microfluidic Device for the Characterisation of Embolisation with Polyvinyl Alcohol Beads through Biomimetic Bifurcations},
  journal      = {Biomedical Microdevices},
  year         = {2012},
  volume       = {14},
  pages        = {153--163},
  doi          = {10.1007/s10544-011-9593-8}
}

@article{p:laurent2004,
  author       = {Laurent, A. and Wassef, M. and Chapot, R. and Houdart, E. and Merland, J.-J.},
  title        = {Location of Vessel Occlusion of Calibrated Tris-Acryl Gelatin Microspheres for Tumor and Arteriovenous Malformation Embolization},
  journal      = {Journal of Vascular and Interventional Radiology},
  year         = {2004},
  volume       = {15},
  number       = {5},
  pages        = {491--496},
  doi          = {10.1097/01.RVI.0000124952.24134.8B}
}

@article{p:osuga2012,
  author       = {Osuga, K. and Maeda, N. and Higashihara, H. and Hori, S. and Nakazawa, T. and Tanaka, K. and Nakamura, M. and Kishimoto, K. and Ono, Y. and Tomiyama, N.},
  title        = {Current Status of Embolic Agents for Liver Tumor Embolization},
  journal      = {International Journal of Clinical Oncology},
  year         = {2012},
  volume       = {17},
  number       = {4},
  pages        = {306--315},
  doi          = {10.1007/s10147-012-0445-1}
}

@article{p:freund2014,
  author       = {Freund, J. B.},
  title        = {Numerical Simulation of Flowing Blood Cells},
  journal      = {Annual Review of Fluid Mechanics},
  year         = {2014},
  volume       = {46},
  pages        = {67--95},
  doi          = {10.1146/annurev-fluid-010313-141349}
}

@article{p:masoud2012,
  author       = {Masoud, H. and Alexeev, A.},
  title        = {Controlled Release of Nanoparticles and Macromolecules from Responsive Microgel Capsules},
  journal      = {ACS Nano},
  year         = {2012},
  volume       = {6},
  number       = {3},
  pages        = {212--219},
  doi          = {10.1021/nn2043143}
}

@article{p:peppas1997,
  author       = {Peppas, N. A.},
  title        = {Hydrogels and Drug Delivery},
  journal      = {Current Opinion in Colloid \& Interface Science},
  year         = {1997},
  volume       = {2},
  number       = {5},
  pages        = {531--537},
  doi          = {10.1016/S1359-0294(97)80103-3}
}

@article{p:hoare2008,
  author       = {Hoare, T. R. and Kohane, D. S.},
  title        = {Hydrogels in Drug Delivery: Progress and Challenges},
  journal      = {Polymer},
  year         = {2008},
  volume       = {49},
  number       = {8},
  pages        = {1993--2007},
  doi          = {10.1016/j.polymer.2008.01.027}
}

@article{p:she2012,
  author       = {She, S. and Xu, C. and Yin, X. and Tong, W. and Gao, C.},
  title        = {Shape Deformation and Recovery of Multilayer Microcapsules after Being Squeezed through a Microchannel},
  journal      = {Langmuir},
  year         = {2012},
  volume       = {28},
  number       = {12},
  pages        = {5010--5016},
  doi          = {10.1021/la3003299}
}

@article{p:lei2020,
  author       = {Lei, W. and Liu, T. and Xie, C. and Yang, H. and Wu, T. and Wang, M.},
  title        = {Enhanced Oil Recovery Mechanism and Recovery Performance of Micro-Gel Particle Suspensions by Microfluidic Experiments},
  journal      = {Energy Science \& Engineering},
  year         = {2020},
  volume       = {8},
  number       = {3},
  pages        = {986--995},
  doi          = {10.1002/ese3.563}
}

@article{p:shi2011,
  author       = {Shi, J. and Varavei, A. and Huh, C. and Delshad, M. and Sepehrnoori, K. and Li, X.},
  title        = {Transport Model Implementation and Simulation of Microgel Processes for Conformance and Mobility Control Purposes},
  journal      = {Energy \& Fuels},
  year         = {2011},
  volume       = {25},
  number       = {11},
  pages        = {5063--5075},
  doi          = {10.1021/ef200835c}
}

@article{p:son2016,
  author       = {Son, H. A. and Choi, S. K. and Jeong, E. S. and Kim, B. and Kim, H. T. and Sung, W. M. and Kim, J. W.},
  title        = {Microbial Activation of Bacillus subtilis-Immobilized Microgel Particles for Enhanced Oil Recovery},
  journal      = {Langmuir},
  year         = {2016},
  volume       = {32},
  number       = {35},
  pages        = {8909--8915},
  doi          = {10.1021/acs.langmuir.6b02010}
}

@article{p:merker2011,
  author       = {Merkel, T. J. and Jones, S. W. and Herlihy, K. P. and Kersey, F. R. and Shields, A. R. and Napier, M. and Luft, J. C. and Wu, H. and Zamboni, W. C. and Wang, A. Z. and Bear, J. E. and DeSimone, J. M.},
  title        = {Using Mechanobiological Mimicry of Red Blood Cells to Extend Circulation Times of Hydrogel Microparticles},
  journal      = {Proceedings of the National Academy of Sciences},
  year         = {2011},
  volume       = {108},
  number       = {2},
  pages        = {586--591},
  doi          = {10.1073/pnas.1010013108}
}

@article{p:li2015,
  author       = {Li, Y. and Sar{\i}yer, O. S. and Ramachandran, A. and Panyukov, S. and Rubinstein, M. and Kumacheva, E.},
  title        = {Universal Behavior of Hydrogels Confined to Narrow Capillaries},
  journal      = {Scientific Reports},
  year         = {2015},
  volume       = {5},
  pages        = {17017},
  doi          = {10.1038/srep17017}
}

@article{p:li2017,
  author       = {Li, Y. and Pan, C. and Li, Y. and Kumacheva, E. and Ramachandran, A.},
  title        = {An Exploration of the Reflow Technique for the Fabrication of an in Vitro Microvascular System to Study Occlusive Clots},
  journal      = {Biomedical Microdevices},
  year         = {2017},
  volume       = {19},
  number       = {4},
  pages        = {82},
  doi          = {10.1007/s10544-017-0213-0}
}

@article{p:brooks2020,
  author       = {Brooks, J. and Minnick, G. and Mukherjee, P. and Jaberi, A. and Chang, L. and Espinosa, H. D. and Yang, R.},
  title        = {High Throughput and Highly Controllable Methods for In Vitro Intracellular Delivery},
  journal      = {Small},
  year         = {2020},
  volume       = {16},
  number       = {47},
  pages        = {2004917},
  doi          = {10.1002/smll.202004917}
}

@article{p:aguiar2019,
  author       = {Bouhid de Aguiar, I. and Meireles, M. and Bouchoux, A. and Schro{\"e}n, K.},
  title        = {Microfluidic Model Systems Used to Emulate Processes Occurring during Soft Particle Filtration},
  journal      = {Scientific Reports},
  year         = {2019},
  volume       = {9},
  pages        = {3063},
  doi          = {10.1038/s41598-019-39820-z}
}

@article{p:lei2019,
  author       = {Lei, W. and Xie, C. and Wu, T. and Wu, X. and Wang, M.},
  title        = {Transport Mechanism of Deformable Micro-Gel Particle through Micropores with Mechanical Properties Characterized by AFM},
  journal      = {Scientific Reports},
  year         = {2019},
  volume       = {9},
  pages        = {1453},
  doi          = {10.1038/s41598-018-37270-7}
}

@article{p:connell2019,
  author       = {O'Connell, M. G. and Lu, N. B. and Browne, C. A. and Datta, S. S.},
  title        = {Cooperative Size Sorting of Deformable Particles in Porous Media},
  journal      = {Soft Matter},
  year         = {2019},
  volume       = {15},
  number       = {18},
  pages        = {3620--3627},
  doi          = {10.1039/C9SM00300B}
}

@article{p:khan2017,
  author       = {Khan, Z. S. and Kamyabi, N. and Hussain, F. and Vanapalli, S. A.},
  title        = {Passage Times and Friction Due to Flow of Confined Cancer Cells, Drops, and Deformable Particles in a Microfluidic Channel},
  journal      = {Convergent Science Physical Oncology},
  year         = {2017},
  volume       = {3},
  number       = {2},
  pages        = {024001},
  doi          = {10.1088/2057-1739/aa5f60}
}

@article{p:fai2017,
  author       = {Fai, T. G. and Kusters, R. and Harting, J. and Rycroft, C. H. and Mahadevan, L.},
  title        = {Active Elastohydrodynamics of Vesicles in Narrow Blind Constrictions},
  journal      = {Physical Review Fluids},
  year         = {2017},
  volume       = {2},
  number       = {11},
  pages        = {113601},
  doi          = {10.1103/PhysRevFluids.2.113601}
}

@article{p:nascimento2017,
  author       = {do Nascimento, D. F. and Avenda{\~n}o, J. A. and Mehl, A. and Moura, M. J. B. and Carvalho, M. S. and Duncanson, W. J.},
  title        = {Flow of Tunable Elastic Microcapsules through Constrictions},
  journal      = {Scientific Reports},
  year         = {2017},
  volume       = {7},
  pages        = {11898},
  doi          = {10.1038/s41598-017-11950-2}
}

@article{p:portnov2018,
  author       = {Portnov, I. V. and M{\"o}ller, M. and Richtering, W. and Potemkin, I. I.},
  title        = {Microgel in a Pore: Intraparticle Segregation or Snail-like Behavior Caused by Collapse and Swelling},
  journal      = {Macromolecules},
  year         = {2018},
  volume       = {51},
  number       = {21},
  pages        = {8147--8157},
  doi          = {10.1021/acs.macromol.8b01569}
}

@article{p:zhang2018,
  author       = {Zhang, Z. and Xu, J. and Drapaca, C.},
  title        = {Particle Squeezing in Narrow Confinements},
  journal      = {Microfluidics and Nanofluidics},
  year         = {2018},
  volume       = {22},
  number       = {11},
  pages        = {120},
  doi          = {10.1007/s10404-018-2129-2}
}

@article{p:keith2020,
  author       = {Keith, A. N. and Vatankhah-Varnosfaderani, M. and Clair, C. and Fahimipour, F. and Dashtimoghadam, E. and Lallam, A. and Sztucki, M. and Ivanov, D. A. and Liang, H. and Dobrynin, A. V. and Sheiko, S. S.},
  title        = {Bottlebrush Bridge between Soft Gels and Firm Tissues},
  journal      = {ACS Central Science},
  year         = {2020},
  volume       = {6},
  number       = {3},
  pages        = {413--419},
  doi          = {10.1021/acscentsci.9b01216}
}

@article{p:buning2021,
  author       = {B{\"u}ning, D. and Schumacher, J. and Helling, A. and Chakroun, R. and Ennen-Roth, F. and Gr{\"o}schel, A. H. and Thom, V. and Ulbricht, M.},
  title        = {Soft Synthetic Microgels as Mimics of Mycoplasma},
  journal      = {Soft Matter},
  year         = {2021},
  volume       = {17},
  number       = {22},
  pages        = {6445--6453},
  doi          = {10.1039/D1SM00379H}
}

@article{p:li2020,
  author       = {Li, S. and Yu, H. and Li, T.-D. and Chen, Z. and Deng, W. and Anbari, A. and Fan, J.},
  title        = {Understanding Transport of an Elastic, Spherical Particle through a Confining Channel},
  journal      = {Applied Physics Letters},
  year         = {2020},
  volume       = {116},
  number       = {10},
  pages        = {103705},
  doi          = {10.1063/1.5139887}
}

@article{p:lesage2009,
  author       = {Lesage, D. and Angelini, E. D. and Bloch, I. and Funka-Lea, G.},
  title        = {A Review of 3D Vessel Lumen Segmentation Techniques: Models, Features and Extraction Schemes},
  journal      = {Medical Image Analysis},
  year         = {2009},
  volume       = {13},
  number       = {6},
  pages        = {819--845},
  doi          = {10.1016/j.media.2009.07.011}
}

@article{p:oevreeide2021,
  author       = {Oevreeide, I. H. and Szydlak, R. and Luty, M. and Ahmed, H. and Prot, V. and Skallerud, B. H. and Zem{\l}a, J. and Lekka, M. and Stokke, B. T.},
  title        = {On the Determination of Mechanical Properties of Aqueous Microgels—Towards High-Throughput Characterization},
  journal      = {Gels},
  year         = {2021},
  volume       = {7},
  number       = {2},
  pages        = {64},
  doi          = {10.3390/gels7020064}
}

@article{p:dai2010,
  author       = {Dai, J. and Grace, J. R.},
  title        = {Blockage of Constrictions by Particles in Fluid--Solid Transport},
  journal      = {International Journal of Multiphase Flow},
  year         = {2010},
  volume       = {36},
  number       = {2},
  pages        = {78--87},
  doi          = {10.1016/j.ijmultiphaseflow.2009.08.001}
}

@article{p:rorai2015,
  author       = {Rorai, C. and Touchard, A. and Zhu, L. and Brandt, L.},
  title        = {Motion of an Elastic Capsule in a Constricted Microchannel},
  journal      = {The European Physical Journal E},
  year         = {2015},
  volume       = {38},
  number       = {5},
  pages        = {49},
  doi          = {10.1140/epje/i2015-15049-8}
}

@article{p:kuriakose2013,
  author       = {Kuriakose, S. and Dimitrakopoulos, P.},
  title        = {Deformation of an Elastic Capsule in a Rectangular Microfluidic Channel},
  journal      = {Soft Matter},
  year         = {2013},
  volume       = {9},
  number       = {16},
  pages        = {4284--4296},
  doi          = {10.1039/c3sm27683j}
}

@article{p:fujiyabu2017,
  author       = {Fujiyabu, T. and Li, X. and Shibayama, M. and Chung, U. and Sakai, T.},
  title        = {Permeation of Water through Hydrogels with Controlled Network Structure},
  journal      = {Macromolecules},
  year         = {2017},
  volume       = {50},
  number       = {23},
  pages        = {9411--9416},
  doi          = {10.1021/acs.macromol.7b01807}
}

@article{p:fanucci1988,
  author       = {Fanucci, E. and Orlacchio, A. and Pocek, M.},
  title        = {The vascular geometry of human arterial bifurcations},
  journal      = {Investigative Radiology},
  year         = {1988},
  volume       = {23},
  number       = {10},
  pages        = {713--718},
  doi          = {10.1097/00004424-198810000-00002}
}

@book{b:landau1959,
  author       = {Landau, L. D. and Lifshitz, E. M.},
  title        = {Course of Theoretical Physics, Vol. 6: Fluid Mechanics},
  publisher    = {Pergamon Press},
  year         = {1959}
}

@book{b:batchelor2000,
  author       = {Batchelor, C. K. and Batchelor, G. K.},
  title        = {An Introduction to Fluid Dynamics},
  publisher    = {Cambridge University Press},
  year         = {2000}
}

@article{p:nikoubashman2017,
  author       = {Nikoubashman, A. and Howard, M. P.},
  title        = {Equilibrium Dynamics and Shear Rheology of Semiflexible Polymers in Solution},
  journal      = {Macromolecules},
  year         = {2017},
  volume       = {50},
  number       = {21},
  pages        = {8279--8289},
  doi          = {10.1021/acs.macromol.7b01876}
}

@article{p:milchev2018,
  author       = {Milchev, A. and Egorov, S. A. and Vega, D. A. and Binder, K. and Nikoubashman, A.},
  title        = {Densely Packed Semiflexible Macromolecules in a Rigid Spherical Capsule},
  journal      = {Macromolecules},
  year         = {2018},
  volume       = {51},
  number       = {6},
  pages        = {2002--2015},
  doi          = {10.1021/acs.macromol.7b02643}
}

@article{p:wyss2010,
  author       = {Wyss, H. M. and Franke, T. and Mele, E. and Weitz, D. A.},
  title        = {Capillary Micromechanics: Measuring the Elasticity of Microscopic Soft Objects},
  journal      = {Soft Matter},
  year         = {2010},
  volume       = {6},
  number       = {18},
  pages        = {4550--4555},
  doi          = {10.1039/c003344h}
}

@article{p:laurent2005,
  author       = {Laurent, A. and Wassef, M. and Chapot, R. and Wang, Y. and Houdart, E. and Feng, L. and Huy, P. T. B. and Merland, J.-J.},
  title        = {Partition of Calibrated Tris-Acryl Gelatin Microspheres in the Arterial Vasculature of Embolized Nasopharyngeal Angiofibromas and Paragangliomas},
  journal      = {Journal of Vascular and Interventional Radiology},
  year         = {2005},
  volume       = {16},
  number       = {4},
  pages        = {507--513},
  doi          = {10.1097/01.RVI.0000150038.99488.01}
}

@article{p:stukowski2014,
  author       = {Stukowski, A.},
  title        = {Computational Analysis Methods in Atomistic Modeling of Crystals},
  journal      = {JOM},
  year         = {2014},
  volume       = {66},
  number       = {3},
  pages        = {399--407},
  doi          = {10.1007/s11837-013-0827-5}
}

@book{b:Muthukumar2011,
  author    = {Muthukumar, M.},
  title     = {Polymer Translocation},
  publisher = {CRC Press},
  year      = {2011},
  doi       = {10.1201/b10901}
}

@article{p:Muthukumar2016,
  author  = {Shojaei, Hamid R. and Muthukumar, Murugappan},
  title   = {Translocation of an Incompressible Vesicle through a Pore},
  journal = {J. Phys. Chem. B},
  volume  = {120},
  pages   = {6102--6109},
  year    = {2016},
  doi     = {10.1021/acs.jpcb.6b02079},
  url     = {https://doi.org/10.1021/acs.jpcb.6b02079}
}

@article{p:Doi2018,
  author  = {Khunpetch, Petch and Man, Xingkun and Kawakatsu, Toshihiro and Doi, Masao},
  title   = {Translocation of a vesicle through a narrow hole across a membrane},
  journal = {J. Chem. Phys.},
  volume  = {148},
  pages   = {134901},
  year    = {2018},
  doi     = {10.1063/1.5013677},
  url     = {https://doi.org/10.1063/1.5013677}
}

@article{p:Shi2019,
  author  = {Han, Yunlong and Lin, Hao and Ding, Mingming and Li, Rui and Shi, Tongfei},
  title   = {Flow-induced translocation of vesicles through a narrow pore},
  journal = {Soft Matter},
  volume  = {15},
  pages   = {3307--3314},
  year    = {2019},
  doi     = {10.1039/c9sm00116f},
  url     = {https://doi.org/10.1039/c9sm00116f}
}

@article{p:Milchev2022,
  author  = {Ranguelov, Bogdan and Milchev, Andrey},
  title   = {Translocation kinetics of vesicles through narrow pores},
  journal = {EPL},
  volume  = {138},
  pages   = {42001},
  year    = {2022},
  doi     = {10.1209/0295-5075/ac6c07},
  url     = {https://doi.org/10.1209/0295-5075/ac6c07}
}

@article{p:Doi2023,
  author  = {Zheng, Bin and Ye, Fangfu and Komura, Shigeyuki and Doi, Masao},
  title   = {Universality in the Dynamics of Vesicle Translocation through a Hole},
  journal = {Langmuir},
  volume  = {39},
  pages   = {563--569},
  year    = {2023},
  doi     = {10.1021/acs.langmuir.2c02835},
  url     = {https://doi.org/10.1021/acs.langmuir.2c02835}
}

@article{p:Auth2025,
  author  = {Baruah, Nishant and Midya, Jiarul and Gompper, Gerhard and Dasanna, Anil Kumar and Auth, Thorsten},
  title   = {Adhesion-driven vesicle translocation through membrane-covered pores},
  journal = {Biophys. J.},
  volume  = {124},
  pages   = {740--752},
  year    = {2025},
  doi     = {10.1016/j.bpj.2025.01.012},
  url     = {https://doi.org/10.1016/j.bpj.2025.01.012}
}

@article{p:Peng2023,
  author  = {Borbas, S. W. and Shen, K. and Ji, C. and Viallat, A. and Helfer, E. and Peng, Z.},
  title   = {Transit Time Theory for a Droplet Passing through a Slit in Pressure-Driven Low Reynolds Number Flows},
  journal = {Micromachines},
  year    = {2023},
  volume  = {14},
  number  = {11},
  pages   = {2040},
  doi     = {10.3390/mi14112040}
}

@article{p:Peng2023b,
  author  = {Tang, Zhengxin and Yaya, Fran{\c{c}}ois and Sun, Ethan T. and Shah, Lubna and Xu, Jie and Viallat, Annie and Helfer, Emmanu{\`e}le and Peng, Zhangli},
  title   = {Analytical theory for a droplet squeezing through a circular pore in creeping flows under constant pressures},
  journal = {Physics of Fluids},
  year    = {2023},
  volume  = {35},
  number  = {8},
  pages   = {082016},
  doi     = {10.1063/5.0156349}
}

@article{p:Park1996,
  author  = {Sung, W. and Park, P. J.},
  title   = {Polymer Translocation through a Pore in a Membrane},
  journal = {Phys. Rev. Lett.},
  volume  = {77},
  pages   = {783--786},
  year    = {1996},
  doi     = {10.1103/PhysRevLett.77.783}
}

@article{p:Nelson1999,
  author  = {Lubensky, D. K. and Nelson, D. R.},
  title   = {Driven Polymer Translocation through a Narrow Pore},
  journal = {Biophys. J.},
  volume  = {77},
  pages   = {1824--1838},
  year    = {1999},
  doi     = {10.1016/S0006-3495(99)77043-6}
}

@article{p:Kardar2002,
  author  = {Chuang, Jeffrey and Kantor, Yacov and Kardar, Mehran},
  title   = {Anomalous dynamics of translocation},
  journal = {Physical Review E},
  year    = {2002},
  volume  = {65},
  number  = {1},
  pages   = {011802},
  doi     = {10.1103/PhysRevE.65.011802}
}

@article{p:Kardar2004,
  author  = {Kantor, Yacov and Kardar, Mehran},
  title   = {Anomalous dynamics of forced translocation},
  journal = {Physical Review E},
  year    = {2004},
  volume  = {69},
  number  = {2},
  pages   = {021806},
  doi     = {10.1103/PhysRevE.69.021806}
}

@article{p:Sakaue2007,
  author  = {Sakaue, T.},
  title   = {Nonequilibrium Dynamics of Polymer Translocation and Straightening},
  journal = {Phys. Rev. E},
  volume  = {76},
  pages   = {021803},
  year    = {2007},
  doi     = {10.1103/PhysRevE.76.021803}
}

@article{p:Aksimentiev2004,
  author  = {Aksimentiev, Aleksij and Heng, Jiunn B. and Timp, Gregory and Schulten, Klaus},
  title   = {Microscopic kinetics of {DNA} translocation through synthetic nanopores},
  journal = {Biophysical Journal},
  year    = {2004},
  volume  = {87},
  number  = {3},
  pages   = {2086--2097},
  doi     = {10.1529/biophysj.104.042960}
}

@article{p:Timp2005,
  author  = {Heng, Jiunn B. and Aksimentiev, Aleksei and Ho, Chuen and Marks, Patrick and Grinkova, Yelena V. and Sligar, Steve and Schulten, Klaus and Timp, Gregory},
  title   = {Stretching {DNA} Using the Electric Field in a Synthetic Nanopore},
  journal = {Nano Letters},
  year    = {2005},
  volume  = {5},
  number  = {10},
  pages   = {1883--1888},
  doi     = {10.1021/nl0510816}
}

@article{r:Palyulin2014,
  author  = {Palyulin, Vladimir V. and Ala-Nissila, Tapio and Metzler, Ralf},
  title   = {Polymer translocation: the first two decades and the recent diversification},
  journal = {Soft Matter},
  year    = {2014},
  volume  = {10},
  pages   = {9016--9037},
  doi     = {10.1039/c4sm01819b}
}

@article{p:Sakaue2016,
  author  = {Sakaue, Takahiro},
  title   = {Dynamics of Polymer Translocation: A Short Review with an Introduction of Weakly-Driven Regime},
  journal = {Polymers},
  year    = {2016},
  volume  = {8},
  number  = {12},
  pages   = {424},
  doi     = {10.3390/polym8120424}
}

@article{r:Wanunu2012,
  author  = {Wanunu, M.},
  title   = {Nanopores: A Journey towards DNA Sequencing},
  journal = {Phys. Life Rev.},
  volume  = {9},
  pages   = {125--158},
  year    = {2012},
  doi     = {10.1016/j.plrev.2012.05.010}
}

@article{p:Baroud2023,
  author = {Moore, Charles Paul and Husson, Julien and Boudaoud, Arezki and Amselem, Gabriel and Baroud, Charles N.},
  title = {Clogging of a {Rectangular} {Slit} by a {Spherical} {Soft} {Particle}},
  journal = {Physical Review Letters},
  volume = {130},
  number = {6},
  pages = {064001},
  year = {2023},
  month = {feb},
  publisher = {American Physical Society},
  doi = {10.1103/PhysRevLett.130.064001}
  }

@article{p:Peng2023c,
  title={Physical mechanisms of red blood cell splenic filtration},
  author={Moreau, Alexis and Yaya, Fran{\c{c}}ois and Lu, Huijie and Surendranath, Anagha and Charrier, Anne and Dehapiot, Benoit and Helfer, Emmanu{\`e}le and Viallat, Annie and Peng, Zhangli},
  journal={Proceedings of the National Academy of Sciences},
  volume={120},
  number={44},
  pages={e2300095120},
  year={2023},
  publisher={National Academy of Sciences},
  doi={10.1073/pnas.2300095120}
  }

@article{p:Guck2024,
  author  = {Reichel, Felix and Goswami, Ruchi and Girardo, Salvatore and Guck, Jochen},
  title   = {High-throughput viscoelastic characterization of cells in hyperbolic microchannels},
  journal = {Lab on a Chip},
  year    = {2024},
  volume  = {24},
  pages   = {2440--2453},
  doi     = {10.1039/d3lc01061a}
}

@article{p:Viallat2020,
  author  = {Dupire, J. and Puech, P.-H. and Helfer, E. and Viallat, A.},
  title   = {Mechanical adaptation of monocytes in model lung capillary networks},
  journal = {Proceedings of the National Academy of Sciences of the United States of America},
  year    = {2020},
  volume  = {117},
  number  = {25},
  pages   = {14798--14804},
  doi     = {10.1073/pnas.1919984117}
}

@article{p:Meller2008,
  title={Orientation-dependent interactions of {DNA} with an $\alpha$-hemolysin channel},
  author={Wanunu, Meni and Chakrabarti, Buddhapriya and Math{\'e}, J{\'e}r{\^o}me and Nelson, David R. and Meller, Amit},
  journal={Physical Review E},
  volume={77},
  number={3},
  pages={031904},
  year={2008},
  publisher={APS},
  doi={10.1103/PhysRevE.77.031904}
}

@article{p:Yodh2012,
  author  = {Wen, Qi and Basu, Anindita and Janmey, Paul A. and Yodh, Arjun G.},
  title   = {Non-affine deformations in polymer hydrogels},
  journal = {Soft Matter},
  year    = {2012},
  volume  = {8},
  pages   = {8039--8049},
  doi     = {10.1039/c2sm25364j}
}

@article{p:DiDonna2005,
  author  = {DiDonna, B. A. and Lubensky, T. C.},
  title   = {Nonaffine correlations in random elastic media},
  journal = {Physical Review E},
  year    = {2005},
  volume  = {72},
  number  = {6},
  pages   = {066619},
  doi     = {10.1103/PhysRevE.72.066619}
}

@article{p:Mackintosh2003,
  author  = {Head, David A. and Levine, Alex J. and MacKintosh, F. C.},
  title   = {Deformation of Cross-Linked Semiflexible Polymer Networks},
  journal = {Physical Review Letters},
  year    = {2003},
  volume  = {91},
  number  = {10},
  pages   = {108102},
  doi     = {10.1103/PhysRevLett.91.108102}
}

@phdthesis{t:Biswas2022,
  title={Equilibrium, Structural, and Mechanical Properties of Soft and Bio Molecular Systems: from Single Polymer to Polymeric Gels},
  author={Biswas, Subhadip},
  year={2022},
  school={University of Sheffield},
  type={PhD Thesis},
  url={https://etheses.whiterose.ac.uk/id/eprint/30490/}
}

@article{p:Grest1986,
  author  = {Grest, Gary S. and Kremer, Kurt},
  title   = {Molecular dynamics simulation for polymers in the presence of a heat bath},
  journal = {Physical Review A},
  year    = {1986},
  volume  = {33},
  pages   = {3628--3631},
  doi     = {10.1103/PhysRevA.33.3628}
}

@article{milani2026rheofluidics,
  title={Rheofluidics: frequency-dependent rheology of single drops},
  author={Milani, Matteo and Wang, Wenyun and Russotto, Lorenzo and Zong, Weiyu and Jahnke, Kevin and Weitz, David A and Aime, Stefano},
  journal={arXiv preprint arXiv:2601.07461},
  year={2026},
  url={https://doi.org/10.48550/arXiv.2601.07461}
}

@article{milani2026decoupling,
  title = {Decoupling Structure and Elasticity in Colloidal Gels Under Isotropic Compression},
  author = {Milani, M. and Cavalletti, E. and Ruzzi, V. and Martinelli, A. and Dieudonn\'e-George, P. and Ligoure, C. and Phou, T. and Cipelletti, L. and Ramos, L.},
  journal = {Phys. Rev. Lett.},
  volume = {136},
  issue = {3},
  pages = {038201},
  numpages = {7},
  year = {2026},
  month = {Jan},
  publisher = {American Physical Society},
  doi = {10.1103/kjks-x7n9},
  url = {https://link.aps.org/doi/10.1103/kjks-x7n9}
}

@article{p:Kremer1990,
  author  = {Kremer, Kurt and Grest, Gary S.},
  title   = {Dynamics of entangled linear polymer melts: A molecular-dynamics simulation},
  journal = {The Journal of Chemical Physics},
  year    = {1990},
  volume  = {92},
  number  = {8},
  pages   = {5057--5086},
  doi     = {10.1063/1.458541}
}

@article{p:Andersen1971,
  author  = {Weeks, J. D. and Chandler, D. and Andersen, H. C.},
  title   = {Role of repulsive forces in determining the equilibrium structure of simple liquids},
  journal = {The Journal of Chemical Physics},
  year    = {1971},
  volume  = {54},
  pages   = {5237--5247},
  doi     = {10.1063/1.1674820}
}
\end{document}